\documentclass[twocolumn]{aastex63}
\usepackage{subfigure}
\usepackage{color}
\usepackage{xcolor}
\usepackage{amsmath}
\usepackage{mathrsfs}
\usepackage{multirow}
\usepackage{soul}
\usepackage{amssymb}
\usepackage{hyperref}
\definecolor{ultramarine}{rgb}{0.8, 0.1, 0.4}

\def\be{\begin{equation}}
\def\ee{\end{equation}}
\def\bd{\begin{displaymath}}
\def\ed{\end{displaymath}}
\def\ba{\begin{aligned}}
\def\ea{\end{aligned}}

\def\nms{\mathsurround=0pt}

\def\oversim#1#2{\lower 4pt\vbox{\baselineskip 0pt \lineskip 1pt
    \ialign{$\nms#1\hfil##\hfil$\crcr#2\crcr\sim\crcr}}}
\def\ga{\mathrel{\mathpalette\oversim>}}
\def\la{\mathrel{\mathpalette\oversim<}}

\def\bh{M_{\bullet}}

\def\msun{M_{\odot}}
\def\AU{{\rm AU}}
\def\kms{{\rm km\,s^{-1}}}

\begin{document}

\title{The Eccentric and Accelerating Stellar Binary Black Hole Mergers in Galactic Nuclei: Observing in Ground and Space Gravitational Wave Observatories}
\author{Fupeng Zhang}
\correspondingauthor{FUPENG ZHANG}
\affiliation{School of Physics and Materials Science, Guangzhou
University, Guangzhou 510006, China}[0]
\email{zhangfupeng@gzhu.edu.cn}
\affiliation{Key Laboratory for Astronomical Observation and Technology of Guangzhou, 510006 Guangzhou, China}
\affiliation{Astronomy Science and Technology Research Laboratory of Department of Education of Guangdong Province, Guangzhou 510006, China}
\author{Xian Chen}
\affiliation{Kavli Institute for Astronomy and Astrophysics, Peking University,
Beijing 100871, China}
\affiliation{Astronomy Department, School of Physics, Peking University, Beijing 100871, China}
\author{Lijing Shao}
\affiliation{Kavli Institute for Astronomy and Astrophysics, Peking University,
Beijing 100871, China}
\affiliation{National Astronomical Observatories, Chinese Academy of Sciences, Beijing 100012, China}
\author{Kohei Inayoshi}
\affiliation{Kavli Institute for Astronomy and Astrophysics, Peking University,
Beijing 100871, China}

%\author{FUPENG ZHANG$^{1,5,6}$, Xian Chen$^{2,3}$, Lijing Shao$^{2,4}$, Kohei Inayoshi$^2$}
%\correspondingauthor{FUPENG ZHANG}
%\affil{$^1$\,School of Physics and Materials Science, Guangzhou
%University, Guangzhou 510006, China, zhangfupeng@gzhu.edu.cn\\
%$^2$\,Kavli Institute for Astronomy and Astrophysics, Peking University,
%Beijing 100871, China \\
%$^3$\,Astronomy Department, School of Physics, Peking University, Beijing 100871, P. R. China\\
%$^4$\,National Astronomical Observatories, Chinese Academy of Sciences, Beijing 100012, China\\
%$^5$\,Key Laboratory for Astronomical Observation and Technology of Guangzhou, 510006 Guangzhou, China\\
%%
%$^6$\,Astronomy Science and Technology Research Laboratory of Department of Education of Guangdong Province, Guangzhou 510006, China\\
%}

\begin{abstract}
We study the stellar binary black holes (BBHs) inspiralling/merging in galactic nuclei based on { our numerical method {\texttt{GNC}}}. We find that $3-40\%$ of all new born BBHs will finally merge due to various dynamical effects. In a five year's mission, up to $10^4$, $10^5$, $\sim100$ of BBHs inspiralling/merging in galactic nuclei can be detected with SNR$>8$ in aLIGO, Einstein{/DECIGO}, TianQin/LISA/TaiJi, respectively. About tens are detectable in both LISA/TaiJi/TianQin and aLIGO. These BBHs have two unique characteristics: (1) Significant eccentricities. $1-3\%$, $2-7\%$, or $30-90\%$ of them is with $e_i>0.1$ when they enter into aLIGO, Einstein, or { space observatories}, respectively. 
{ Such high eccentricities provide a possible explanation for that of GW 190521. 
Most highly-eccentric BBHs are not detectable in LISA/Tianqin/TaiJi
before entering into aLIGO/Einstein as their strain become significant 
only at $f_{\rm GW}\ga0.1$ Hz. DECIGO become an ideal observatory to detect those events as it 
can fully cover the rising phase.} (2) Up to $2\%$ of BBHs can inspiral/merge at distances $\la10^3 r_{\rm SW}$ { from the massive black hole (MBH)}, with significant accelerations, such that the Doppler phase drift of $\sim10-10^5$ of them can be detectable with SNR$>8$ in space observatories. The energy density of the gravitational wave backgrounds (GWB) contributed by these BBHs deviate from the powerlaw slope of $2/3$ at 
$f_{\rm GW}\la 1$mHz. The high eccentricity, significant accelerations and different profile of GWB of these sources make them distinguishable, thus interesting for future GW detections and tests of relativities. 
\end{abstract}


\section{Introduction}
The discovery of the merging of binary black holes (BBHs) by the Advanced LIGO (aLIGO) in September 2015
opens a new era of gravitational wave astronomy~\citep{Abbott16a}. After a successive of observational runs 
of aLIGO and Virgo, up to {$50$} compact binary mergers have been revealed~\citep{2021PhRvX..11b1053A,
2019PhRvX...9c1040A, Abbott19,Abbott16b,Abbott16c,Abbott16d,Abbott17a,Abbott17b,Abbott17c,Abbott17d}.
Although the BBHs mergers account for most of these sources, their astrophysical origin
remain largely unclear. The BBHs may merge isolatedly in the field of which the progenitors are binary stars 
in the early universe~\citep[e.g.,][]{Dominik15,Belczynski16} or even population III binary stars~\citep[e.g.,][]{2017MNRAS.468.5020I,2020arXiv200911447L}. 
Many other models propose dynamical channels of merging, e.g., through Kozai-Lidov effect~\citep[e.g.,][]{Kozai62, 2017ApJ...834..200M, Wen03, Antonini12, 2019MNRAS.488...47F}, gas-assistant or dynamical hardening in accretion discs~\citep[e.g.,][]{2017ApJ...835..165B,2017MNRAS.464..946S,2018ApJ...866...66M, 2020ApJ...898...25T},
dynamical interactions in dense-stellar environments, e.g., in globular clusters~\citep[e.g.,][]{Rodriguez15,Rodriguez16, Rodriguez16b}, 
and in the galactic nuclei~\citep[e.g.,][]{Antonini12,Antonini16, Zhang19, 2020ApJ...891...47A}. 

The BBHs merging around the massive black hole (MBH) in galactic nuclei are different from those 
in other channels, mainly due to the existence of the MBH. These merging events have some unique features.
For example, the phase shift of the waveform due to acceleration~\citep[e.g.,][]{2017PhRvD..95d4029B, Inayoshi17, 2019MNRAS.488.5665W, 2020PhRvD.101h3028T}, relativistic effects or gravitational effects~\citep[e.g., ][]{2017ApJ...834..200M}, repeated gravitational lensing of waves~\citep{1973JMP....14....1C, 2020PhRvD.101h3031D,1973PhRvD...7.2275L, 2013ApJ...763..122K}, 
binary extreme-mass ratio inspirals (b-EMRIs)~\citep{2018CmPhy...1...53C},
and significant eccentricities, which may be due to Kozai-Lidov~\citep{Wen03,Hoang18,Antonini12, Zhang19} 
, gravitational wave captures~\citep[e.g.,][]{Oleary09, 2018ApJ...860....5G, 2020arXiv201102507G,2015MNRAS.448..754H}
or three-body encounters~\citep[e.g.,][]{Zhang19,2019ApJ...885..135T}.
These features are significantly different from those of BBHs merging in other environments. For example, high eccentricity 
will not be expected in aLIGO/Virgo or LISA band for those BBHs merging isolatedly in the field. Also, the accelerations of 
BBHs merging in the globular clusters are expected several orders of magnitude smaller than those merging near the 
MBH~\citep[e.g.,][]{Inayoshi17}. 

In order to separate various merging channels, it is important to study the distribution of eccentricity, and other 
unique features associated with the BBHs merging around the MBHs that can possibly be measured in the 
current and future gravitational wave (GW) observatories~\citep[e.g.,][]{2019DDA....5020203H, 2019arXiv190208604R, 2019ApJ...887..210F}. As aLIGO and Virgo only detect the signals in the final seconds, many useful information that 
related to the above effects may have been lost. Thus, to probe these effects, it will be necessary to study the characteristics of
them by GW detectors in other bands, e.g., Einstein~\citep{2011CQGra..28i4013H}, or {Deci-Hz detectors such as DECIGO}~\citep[e.g.,][]{2006CQGra..23S.125K, 2021PTEP.2021eA105K, 2019arXiv190811375A,2020MNRAS.496..182L,decihertz18}
or the space GW detectors, e.g.,  LISA~\citep{AmaroSeoane17}, 
 TaiJi~\citep{2020IJMPA..3550075R} or TianQin~\citep{2016CQGra..33c5010L}  
, in addition to the aLIGO/Virgo detection.

In a previous work~\citep[][here after Paper I]{Zhang19}, we have developed a Monte-Carlo numerical simulation 
framework to study the dynamical evolution and merging of BBHs near the MBH in galactic nuclei. Paper I
has for the first time combined various dynamical effects that are important for the 
evolution of BBHs. The rates and the eccentricities of merging events near local universe in 
aLIGO and Virgo band have been only simply 
discussed in Paper I. Here we make some further improvements of the framework and 
then calculate and study more details of the observational features, e.g., the predicted number, eccentricity and the Doppler drift of the inspiraling/merging BBH with significant signal to noise ratio (SNR) in aLIGO/Virgo, Einstein, TianQin, LISA and TaiJi band. 
We focus mainly on the BBH merging events with high eccentricities, or happen very close to the MBH with 
significant phase drifts. Different observatories can cover only a specific range of frequency of
 a BBH event, thus, it will be also interesting to study those BBHs of which the GW frequency 
 can evolve and cross multiple observatories within a 
 short period of time, e.g., $10$ yr. These events are important for localization and multiband observations of 
the sources in future. 
 
This work is organized as following. In Section~\ref{sec:evl_in_cluster} we obtain samples of BBHs that 
are merging in the galactic nuclei in different models, by using an updated numerical framework from our previous paper. 
These samples are output from the numerical method when the inner orbital evolution is dominated by GW orbital decay. 
We also describe the detail model settings and the method used to obtain the inferred merging rates.
In Section~\ref{sec:OB_LISA_LIGO}, we calculate the 
evolution of the GW radiation of BBHs  and use a Monte-Carlo method to obtain samples of inspiralling/merging BBHs within 
a given period of observation time. We also predict the detectable number of BBHs with SNR$>8$ in different observatories. In Section~\ref{sec:ob_ecc} we study the entering eccentricities of BBHs in different observatories. In Section~\ref{sec:relativity_BBHs} we discuss the BBHs inspiralling/merging very close to the MBH and those samples with significant Doppler phase drifts in different observatories. Section~\ref{sec:GWBK} describe the calculation and the results of the gravitational wave backgrounds from the 
BBHs merging around MBH. { Discussion of results in different models are provided in Section~\ref{sec:discussion}. 
The merging events of BBHs presented in this work are all for first generation of BBH mergers due to a number of limitations
in the current MC code. In order to study multiple generation of mergers and also EMRI events, we plan to update the MC code
in the future, with more details shown in Section~\ref{sec:update}. Finally, the conclusions are shown in Section~\ref{sec:conclusion}.}

In this paper, we assume a flat $\Lambda$CDM cosmology with parameters ($h_0, \Omega_{\rm m}, \Omega_\Lambda)=(0.679, 0.306, 0.694)$~\citep{2014A&A...571A..16P}, where $h_0=H_0/100$\,km s$^{-1}$\,Mpc$^{-1}$ with $H_0$ as the Hubble constant, $\Omega_{\rm m}$ and $\Omega_{\Lambda}$ are the fractions of matter and cosmological constant in the local universe, respectively.

\section{Modeling the evolution of BBHs in the Nuclear star cluster}
\label{sec:evl_in_cluster}

In this section, we describe the model used to calculate the evolution of BBHs in the nuclear star cluster and some 
immediate results from the model, with the assumed initial conditions of the cluster
and the BBHs described in Section~\ref{subsec:initial_condition}. The merging rate estimation have some 
uncertainties, and we have covered them in full details in the Section~\ref{subsec:merging_rate}. 
The results of many possible outcome (including merging, tidal disruption and ionization) of the BBHs from our simulations are summarized in Section~\ref{subsec:evl_merging_BBHs}. 

Our method is based on an updated Monte-Carlo numerical method in our previous work (Paper I). 
We name it the Monte-Carlo code for dynamics of Galactic Nuclear star Cluster with a central MBH (abbreviated to {{\texttt{ GNC}}}).
{{\texttt{ GNC}}} have combined various dynamical effects of BBH in 
its evolution, e.g., the two-body relaxation and resonant relaxation~\citep{RT96} of outer-orbit of BBHs, the encounters of the BBH with the background objects, gravitational wave orbital decays of the inner orbit of BBHs,
Kozai-Lidov (KL) oscillation, and the close encounters of BBHs with the central MBH.
We have used the “clone scheme” similar to~\citet{SM78} to increase the number of BBHs in inner regions. 
These clone BBHs are created once a BBH moves inside some boundaries of the inner regions of the cluster, 
such that the total number of samples in the very inner regions will be boosted to a sufficiently large number to reduce the fluctuations of statistics. For more details, see Paper I\footnote{Clone samples generated by the clone scheme have different weights in statistics. Throughout the paper, the weight of each sample 
has been considered in any statistical calculations, and we does not show details of them for simplicity.}. 

Compared to Paper I, we have made some improvements on {{\texttt{ GNC}}}: 
(1) We can now consider multiple mass components of the background stars (or compact objects)
for both the dynamical relaxation and collisions between the BBHs and the background objects. By such improvement, 
we can consider the effects from different background components, e.g., the stars and the clusters of single black holes.
(2) We have added the GW radiation in any three-body dynamical interactions. For example, during the encounters 
between the BBH with a background star, and during the encounters between the BBHs with the central MBH; 
(3) Other minor improvements. For more details of these above improvements, see the Appendix~\ref{apx:method_update};

Note that, if not otherwise specified, 
we adopt the notation of symbols the same as in our previous work~(See Table 1 of Paper I).

\subsection{Models and the initial conditions}
\label{subsec:initial_condition}
\begin{table}%[h]
\caption{Models and initial conditions}
\centering
\begin{tabular}{lccccccccccccc}\hline
Model & $\log$ $\bh$ &  $r_i^{a}$  & IMF$^{b}$ & pdf($e_{1,\rm ini}$)$^{c}$ & $m_\star$ ($\msun$)\\
\hline
MP1  & $5-8$    & $r_h$      & LIGO\_BK & $U(0,0.9)$        & $1$     \\
MP2  & $5-8$    & $0.1r_h$   & LIGO\_BK & $U(0,0.9)$        & $1$     \\
MP3  & $5-8$    & $r_h$      & LIGO\_BK & $e_{1,\rm ini}=0$           & $1$     \\
MP4  & $5-8$    & $r_h$      & LIGO\_BK & $U(0,0.9)$        & $10$    \\
MP5  & $5-8$    & $r_h$      & $\gamma=-2.35$ & $U(0,0.9)$  & $1$     \\
MP6  & $5-8$    & $r_h$      & LIGO\_BK & $U(0,0.9)$        & $1+10$  \\
\hline
%& 0.99 & 45$\arcdeg$ & 180$\arcdeg$  \\ \hline
%
\end{tabular}
\tablecomments{Initial conditions for models. For more details see Section~\ref{subsec:initial_condition}.\\
$^a$ The initial SMA of the outer orbits of BBHs in the cluster. 
$r_h=\bh G/\sigma_h^2$ is the gravitational influence radius. \\
$^b$ The initial mass function of BBHs in the cluster .\\
$^c$ The initial distribution of eccentricity of inner orbit of BBHs. 
}
\label{tab:model}
\end{table}

In order to make reasonable predictions of the merging events of BBHs, 
the uncertainties in the initial conditions of the nuclear star cluster should be covered
in our simulation. Most of the uncertainties  are the mass 
($m_\star$) and the composition (stars and the compact objects) of the background star, 
the birth position of the BBHs ($r_i$ or $a_2$), the initial mass function (IMF) of the 
BBHs, and the details of the BBHs parameters (e.g., the inner eccentricity distribution
of the BBHs when they initially formed). As the prediction of the merging events of 
the BBHs may depend on some of these initial conditions, we explore 
six different models, with the initial conditions listed in Table~\ref{tab:model}, to 
cover these uncertainties.

BBHs are assumed formed at the inner parts $r_i=0.1r_h$ for model MP2, and 
the outer part of the cluster, e.g.,  $r_i=r_h$ for other models. The initial 
orbital semimajor axis (SMA) of the inner orbit of BBHs $a_1$ are assumed 
to follow a log-normal distribution between $0.002\AU< a_1<50\AU$. The upper limit 
of $50\AU$ is set to avoid wide binaries, and an alternative value of the upper 
and lower limit of $a_1$ can be absorbed into the uncertainties in 
the binary fraction $f_{\rm gBBH}$ discussed in Section~\ref{subsec:merging_rate}. 
The initial value of the inner eccentricities of the BBHs are assumed with $e_{1,\rm ini}=0$ 
for model MP3, or uniformly distributed between $0$ and $0.9$ for other models (``$U(0,0.9)$''
in table~\ref{tab:model}).
The IMF of BBHs are assumed to follow a powerlaw given by 
$f(m_A, m_B)\propto m_A^{-2.35}m_B^{-1}$ and $m_A>m_B$ with $m_A+m_B<100\msun$
for model MP5~\citep{Abbott16d}, where $m_A$ and $m_B$ is the primary and secondary mass component of the BBH. 
For other models, their IMF are according to an updated results from
~\citet[][the smoothed broken power-law model, excluding GW190521]{2021ApJ...913L...7A} (``LIGO\_BK''
in table~\ref{tab:model}).

For the background stars, their distribution can be determined by given slope of density
profile ($\alpha_\star$) and the number of objects within the influence radius ($N(<r_h)$).
all models except MP4 and MP6 assume a single background component of stars of $m_\star=1\msun$, 
with a cusp density profile given by $\alpha_\star=7/4$~\citep{BW76} and $N(<r_h)=\bh/m_\star$. 
In MP4, we assume the same density profile except that the background objects are dominated 
by more massive stellar objects (e.g., stellar black holes) with mass of $m_\star=10\msun$
(and $N(<r_h)=\bh/m_\star$). In MP6 we assume that the background objects are combined by two
mass components with masses of $1\msun$ and $10\msun$, which are supposed to be single main-sequence
stars and black holes, respectively. 
Following~\citet{Alexander09}, we assume that for the stars $m_\star=1\msun$, 
$\alpha_\star=1.4$ and $N(<r_h)=\bh/m_\star$; For the black holes
 $m_\bullet=10\msun$, $\alpha_\bullet=2$, and 
$N(<r_h)=0.05\bh/m_\bullet$.

For each model, we perform four simulations with different masses of MBH,
i.e., given by $\bh=10^5\msun,10^6\msun,10^7\msun$, and $10^8\msun$. 
In each simulation, the calculation ends only if the density profile of BBHs 
are converged. For more details of the Monte-Carlo simulation, see Paper I. 

\subsection{Merging Rate estimations}
\label{subsec:merging_rate}

\begin{table*}%[h]
\centering
\caption{Results}
\setlength{\tabcolsep}{1.4mm}

\begin{tabular}{lc|ccc|ccc|cccc|c}\hline
\multirow{2}{*}{Model} & $\log$ $\bh$ &  \multirow{2}{*}{$f_{\rm mrg}^{\rm eMBH}$} & \multirow{2}{*}{$f_{\rm mrg}^{\rm KL}$} & \multirow{2}{*}{$f^{\rm eBK}_{\rm mrg}$} 
& \multirow{2}{*}{$f_{\rm mrg}$}  & \multirow{2}{*}{$f_{\rm td}$} & \multirow{2}{*}{$f_{\rm ion}$}
&$a_2<10^4$ & $r_{p2}<10^4$ & $r<10^4$ & $e_1$ & $\mathcal{R}_0$ \\
& $(\msun)$ & & & & & & &($r_{\rm SW}$) &($r_{\rm SW}$) & ($r_{\rm SW}$) & $>0.999$ & (Gpc$^{-3}$ yr$^{-1}$) \\ 
\hline
      MP1-5&          5&    $15\%$&     $65\%$&     $19\%$&     $20\%$&     $60\%$&    $0.8\%$&    $2.9\%$&     $17\%$&    $3.0\%$&     $68\%$& \multirow{4}{*}{\(\displaystyle \ba7.9\\\sim79\ea\)}    \\
      MP1-6&          6&   $7.1\%$&     $52\%$&     $41\%$&     $29\%$&     $33\%$&    $3.5\%$&    $4.9\%$&     $20\%$&    $4.7\%$&     $58\%$& \\
      MP1-7&          7&   $2.2\%$&     $31\%$&     $67\%$&     $14\%$&    $9.5\%$&    $2.9\%$&    $5.2\%$&     $22\%$&    $5.4\%$&     $48\%$& \\
      MP1-8&          8&   $0.4\%$&    $7.0\%$&     $93\%$&    $9.8\%$&    $7.1\%$&    $2.5\%$&    $4.5\%$&     $21\%$&    $4.3\%$&     $36\%$& \\
 \hline
      MP2-5&          5&    $15\%$&     $67\%$&     $18\%$&     $27\%$&     $63\%$&    $1.1\%$&    $2.0\%$&     $16\%$&    $2.0\%$&     $70\%$&  \multirow{4}{*}{\(\displaystyle \ba20\\\sim201\ea\)} \\
      MP2-6&          6&   $6.1\%$&     $60\%$&     $34\%$&     $43\%$&     $37\%$&    $3.5\%$&    $3.7\%$&     $17\%$&    $3.6\%$&     $64\%$&  \\
      MP2-7&          7&   $2.2\%$&     $45\%$&     $53\%$&     $39\%$&     $13\%$&    $5.3\%$&    $3.4\%$&     $17\%$&    $3.5\%$&     $58\%$&  \\
      MP2-8&          8&   $1.7\%$&     $16\%$&     $83\%$&     $33\%$&     $18\%$&    $7.0\%$&    $2.8\%$&     $29\%$&    $3.3\%$&     $37\%$&  \\
 \hline
      MP3-5&          5&    $19\%$&     $63\%$&     $18\%$&     $19\%$&     $61\%$&    $0.8\%$&    $4.7\%$&     $25\%$&    $4.5\%$&     $66\%$& \multirow{4}{*}{\(\displaystyle \ba7.5\\\sim75\ea\)}      \\
      MP3-6&          6&   $8.0\%$&     $52\%$&     $40\%$&     $28\%$&     $33\%$&    $2.6\%$&    $7.2\%$&     $25\%$&    $6.8\%$&     $57\%$& \\
      MP3-7&          7&   $2.0\%$&     $29\%$&     $69\%$&     $13\%$&    $8.1\%$&    $2.7\%$&    $7.8\%$&     $28\%$&    $7.4\%$&     $44\%$& \\
      MP3-8&          8&   $0.9\%$&    $5.7\%$&     $93\%$&    $8.9\%$&    $7.1\%$&    $2.8\%$&    $6.1\%$&     $26\%$&    $5.9\%$&     $31\%$& \\
 \hline
      MP4-5&          5&    $19\%$&     $51\%$&     $30\%$&    $3.1\%$&     $61\%$&     $10\%$&      $0\%$&     $11\%$&      $0\%$&     $64\%$& \multirow{4}{*}{\(\displaystyle \ba0.2\\\sim1.7\ea\)}    \\
      MP4-6&          6&    $11\%$&     $45\%$&     $44\%$&    $4.3\%$&     $44\%$&     $23\%$&    $0.2\%$&     $10\%$&    $0.2\%$&     $56\%$& \\
      MP4-7&          7&   $6.7\%$&     $28\%$&     $65\%$&    $2.8\%$&     $18\%$&     $18\%$&    $0.5\%$&     $11\%$&    $0.7\%$&     $41\%$& \\
      MP4-8&          8&   $2.5\%$&     $15\%$&     $83\%$&    $2.8\%$&     $14\%$&     $20\%$&    $0.1\%$&    $8.7\%$&    $0.2\%$&     $35\%$& \\
 \hline
      MP5-5&          5&    $13\%$&     $64\%$&     $23\%$&     $16\%$&     $56\%$&    $2.1\%$&    $1.0\%$&     $13\%$&    $1.1\%$&     $69\%$& \multirow{4}{*}{\(\displaystyle \ba5\\\sim50\ea\)}    \\
      MP5-6&          6&   $6.1\%$&     $49\%$&     $45\%$&     $21\%$&     $32\%$&    $6.4\%$&    $3.0\%$&     $16\%$&    $3.1\%$&     $56\%$& \\
      MP5-7&          7&   $2.7\%$&     $26\%$&     $71\%$&    $8.0\%$&    $8.9\%$&    $4.3\%$&    $3.8\%$&     $18\%$&    $3.8\%$&     $47\%$& \\
      MP5-8&          8&   $1.6\%$&    $8.0\%$&     $90\%$&    $6.1\%$&    $7.5\%$&    $3.6\%$&    $3.2\%$&     $15\%$&    $3.0\%$&     $32\%$& \\
 \hline
      MP6-5&          5&    $25\%$&     $68\%$&    $7.0\%$&     $18\%$&     $63\%$&    $5.1\%$&    $6.9\%$&     $25\%$&    $6.7\%$&     $70\%$& \multirow{4}{*}{\(\displaystyle \ba7\\\sim71\ea\)}    \\
      MP6-6&          6&    $11\%$&     $68\%$&     $21\%$&     $25\%$&     $35\%$&     $11\%$&    $9.0\%$&     $27\%$&    $8.2\%$&     $67\%$& \\
      MP6-7&          7&   $4.1\%$&     $44\%$&     $52\%$&     $13\%$&   $10.0\%$&    $7.2\%$&     $11\%$&     $29\%$&     $10\%$&     $56\%$& \\
      MP6-8&          8&   $2.1\%$&     $12\%$&     $86\%$&    $9.6\%$&    $8.8\%$&    $6.0\%$&    $8.7\%$&     $26\%$&    $8.0\%$&     $38\%$&  \\
 \hline
%& 0.99 & 45$\arcdeg$ & 180$\arcdeg$  \\ \hline
%
\end{tabular}
\tablecomments{
$f_{\rm mrg}$, $f_{\rm td}$ and $f_{\rm ion}$ is the fraction of merged, tidal disrupted and 
ionized BBHs in all generated BBHs, respectively. $f_{\rm mrg}^{\rm eMBH}$, $f_{\rm mrg}^{\rm KL}$ and 
 $f^{\rm eBK}_{\rm mrg}$ is the fraction of BBHs merged 
 due to encounters with the central MBH, KL effect, encounters with the background objects, 
 respectively. When the inner orbital evolution of a BBH is dominated by the GW radiation, 
 their orbital parameters, e.g., $a_2$, $r_{p2}$, $r$ and $e_1$ are recorded, 
 where $a_2$ and $r_{p2}$ is the SMA axis and pericenter of the outer orbit of BBHs, respectively,
  $e_1$ is the eccentricity of the inner orbit of BBHs, and $r$ is the distance to the MBH.  $9$-th to $12$-th column show  
the percentage of these parameters above or below some values. 
 $\mathcal{R}_0$ is the estimated total merging rate from local universe, assuming 
constant supply rate of MBHs (See Equation~\ref{eq:RS}). The two numbers 
for each cell show the upper and lower limit, which is obtained according to 
$10^{-4}<f_{\rm gBBH}<{\rm min}(10^{-3}, 0.1 m_\star/\langle m_{\rm BBH}\rangle)$.
}
\label{tab:model_frac}
\end{table*}

One physically motivated method of rate estimation assumes that the supply of BBHs are given by 
the continuous star formation processes that form the 
central nuclear star cluster~\citep[e.g.,][]{Petrovich17,Hamers18}.
We assume that a star cluster of mass $\sim\bh$ is built up in time of about the age of galaxy 
(usually $t_g\sim 10$ Gyr) by continuous star formation, and the fraction of BBHs formed from these new born (binary) stars 
is given by $f_{\rm gBBH}$, then the expected supply rate of BBHs 
is given by $f_{\rm gBBH}\bh/(t_g m_\star)$. During an equilibrium state, if the merge fraction is given by 
$f_{\rm mrg}$ for all new born BBHs, the merging rate of BBHs is given by 
\begin{equation}
\ba
\mathbf{R}(\bh)=&10 {~\rm Gyr^{-1}}\times \frac{f_{\rm mrg}}{0.1} \frac{f_{\rm gBBH}}{10^{-3}}\frac{\bh}{10^6~\msun}
\frac{1~\msun}{m_\star}.
\label{eq:RS}
\ea
\end{equation}

In the above method, $f_{\rm mrg}$ can be obtained from our numerical simulations, 
and $f_{\rm gBBH}$ can be reasonably estimated from stellar evolution models. 
\citet{Belczynski04} simulate the stellar evolution of $10^6$ stellar binaries and found that 
in most cases there will be $\sim 10^3-10^4$ BBHs, suggesting that
a fraction of $10^{-3}-10^{-2}$ of all stellar binaries will end up with BBHs. 
The fraction of binary stars in galactic nuclei can be inferred from the observations of our 
Galactic Center, e.g., $\sim0.3$~\citep{2014ApJ...782..101P}. Thus 
we have $f_{\rm gBBH}\sim 10^{-4}-10^{-3}$.
Considering that currently there are large uncertainties of the mass function of 
BBHs resulting from a given stellar IMF, the above analysis provides only a rough estimation of $f_{\rm gBBH}$. 
To be conservative, we assume that the total masses of newly formed BBHs 
should not exceeds $10\%$ of the all newly formed stellar masses in the cluster (a limit according to~\citet{Alexander09}, in which a canonical IMF of stars is assumed), i.e., $f_{\rm gBBH}\langle m_{\rm BBH}\rangle \la 0.1m_\star$, where $\langle m_{\rm BBH}\rangle$ 
is the average mass of the BBHs. Thus, for all models in Table~\ref{tab:model} we assume that $10^{-4}<f_{\rm gBBH}<{\rm min}
(10^{-3}, 0.1 m_\star/\langle m_{\rm BBH}\rangle)$.

All the above rates are for a galaxy with a given mass of MBH $\bh$. To estimate the total rate for all galaxies in the local universe, 
we integrate it over all the MBHs, 
\be
\mathcal{R}_0=\int_{10^5\msun}^{10^8\msun} \mathbf{R}(\bh)n(\bh, z=0) d\bh
\label{eq:Rate_tot}
\ee

Here the upper limit is $10^8\msun$ as we found that merging events from $>10^8\msun$ are 
negligible~\citep{Zhang19}. Here $n(\bh, z)$ is given by~\citet{2015ApJ...810...74A}, which includes 
both the black hole mass function in inactive and active galaxies. 

\subsection{Evolution and Merging of BBHs}
\label{subsec:evl_merging_BBHs}

%\begin{figure*}
%\center
%\includegraphics[scale=0.45]{fig_track_By_MBH.eps}
%\caption{}
%%
%\label{fig:fbhbmbh}
%\end{figure*}

\begin{figure*}
\center
\includegraphics[scale=0.75]{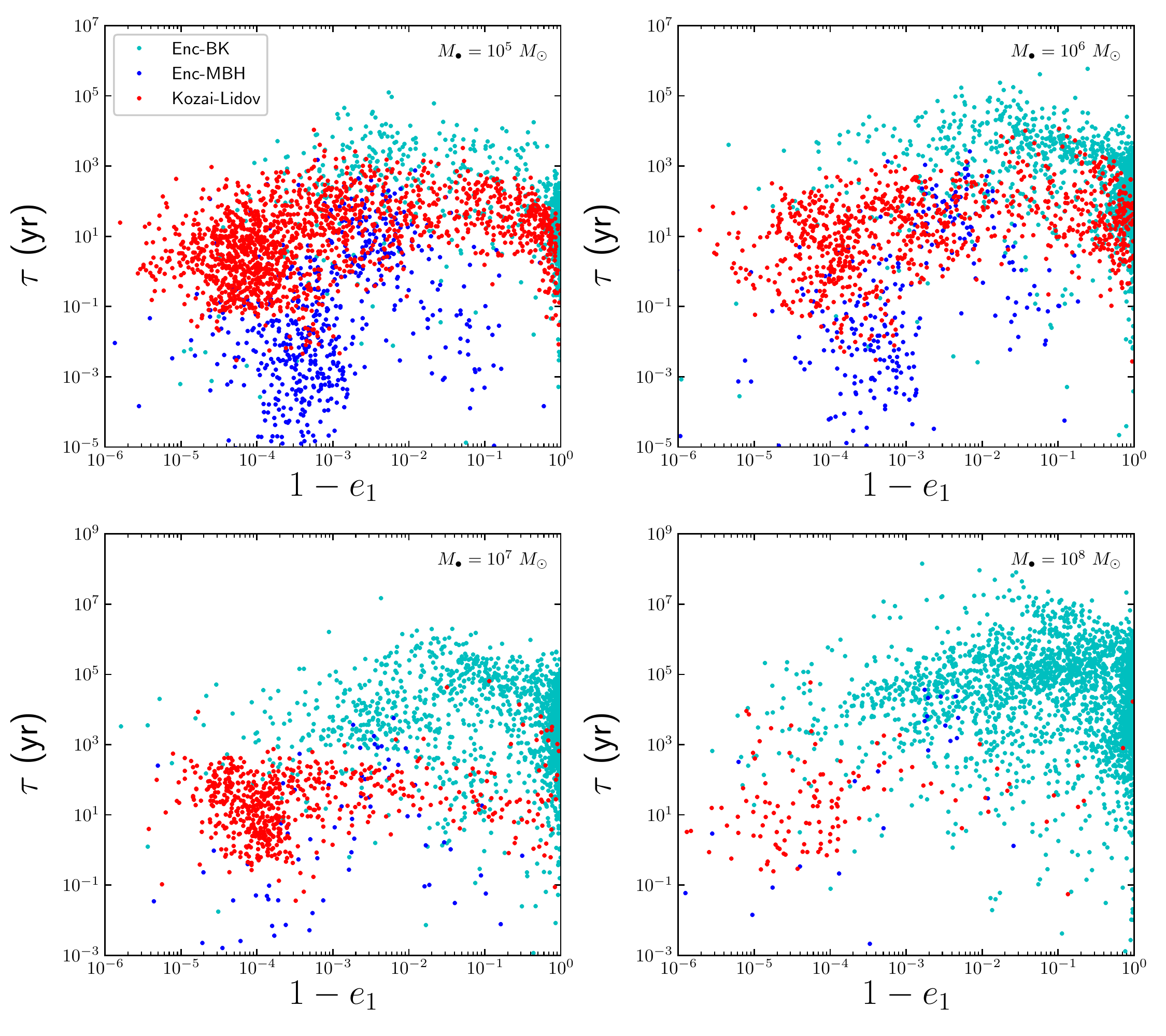}
\caption{The output inner eccentricity and lifetime of the BBHs (inspiralling time from when gravitational wave decay dominates the evolution of the inner orbit to the moment of the final coalescence) of model MP1 given different MBH masses. For clarity, only $3000$ samples are shown in each panel.
 The colored dots show the dynamical process experienced by the BBHs before merging: 
 the KL oscillation (red dots), encounters with a background object (cyan dots) and 
 with the MBH (blue dots).
 }
\label{fig:tgw_e_MP1}
\end{figure*}

\begin{figure*}
\center
\includegraphics[scale=4]{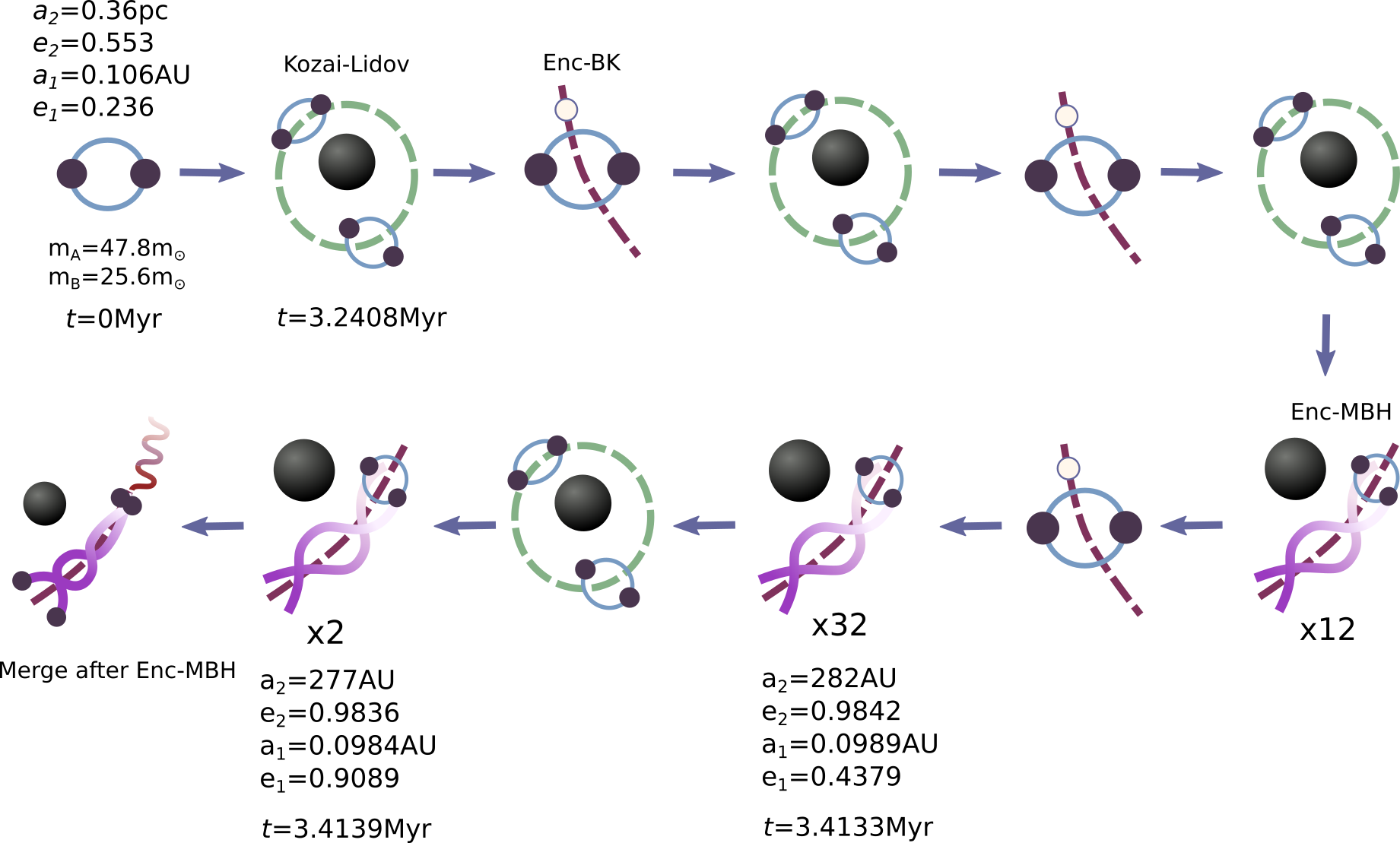}
\includegraphics[scale=0.7]{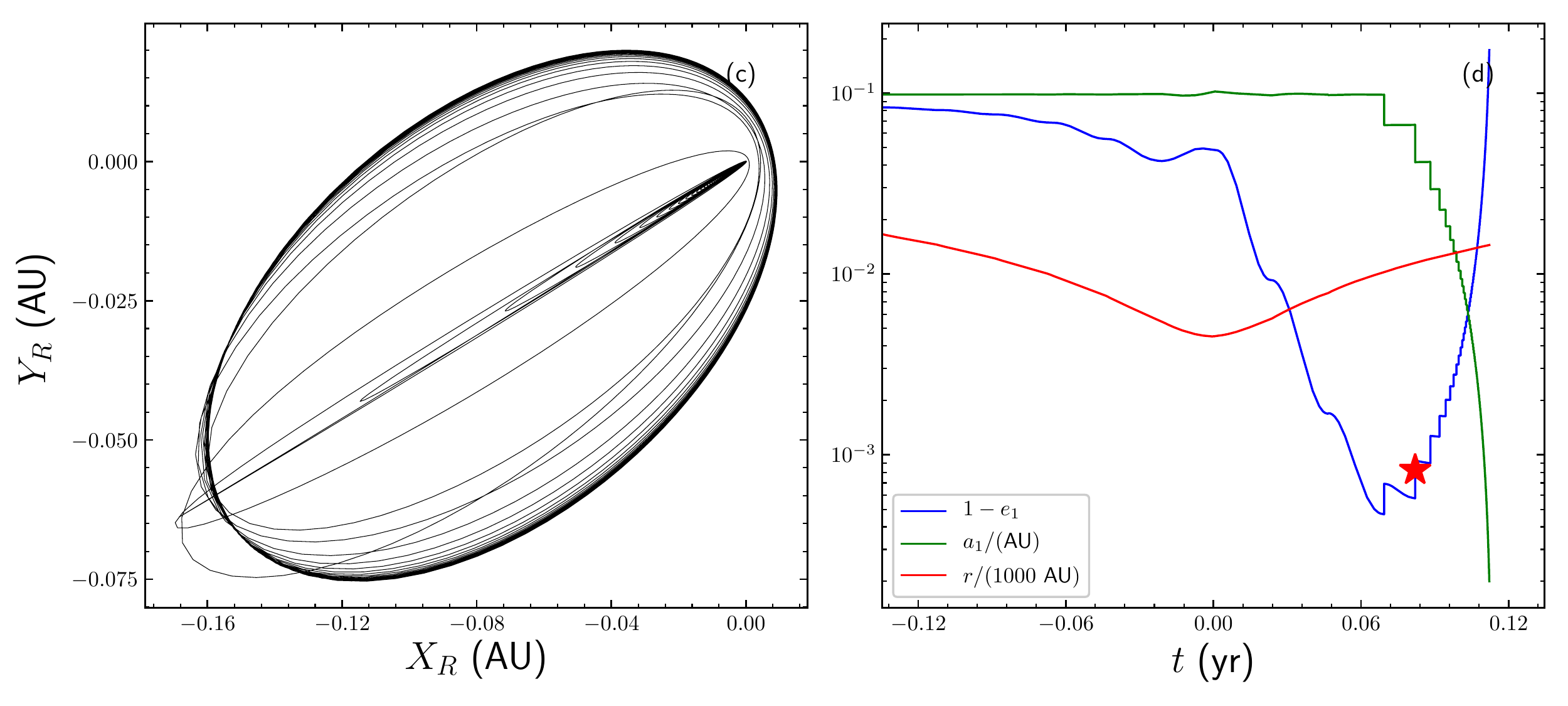}
\caption{Upper illustrations show one example of the dynamical evolution of a BBH that finally merge due to an
encounter with the MBH. The example is from model MP1 of simulation runs given $\bh=10^5\msun$. 
``Kozai-Lidov'', ``Enc-MBH'', and ``Enc-BK'' in the illustration show the dynamical process of KL oscillation, 
 an encounter with the MBH,  and an encounter with a background object, respectively. 
 The trajectory of the relative position vector between the two components of the BBH 
 during the final encounter with the MBH is shown in 
 the bottom left panel. The evolution of the SMA ($a_1$), eccentricity ($e_1$) of the inner orbit of BBH 
 and the distance $(r)$ of the mass center of BBH to the MBH during the final encounter with the MBH are shown in 
 the bottom right panel. The red star in the eccentricity curve marks the 
 position where the {{\texttt{ GNC}}} find that the GW orbital decay dominates the evolution of the inner orbit of the BBH.}
\label{fig:track_fbhbmbh}
\end{figure*}

%{\color{red}The BBHs exit of the program due to different critaria and we are 
%not able to determine the time when the GW starts to dominate the evolution and finally 
%merging. Thus, the samples should have different lifespan from show up with GW dominating and 
%merging. }

In each simulation, BBHs form continuously, and the distribution of BBHs dynamically evolve into an 
equilibrium state when the density profile of BBHs is converged at any position of the cluster. 
During a time interval ${\Delta}t$, suppose that there are a number of ${\Delta}N_{\rm tot}$ BBHs generated in the 
simulation, and a number of ${\Delta}N_{\rm mrg}$ of samples end up with merging, then we can estimate that the 
corresponding merging fractions of BBHs are given by $f_{\rm mrg}={\Delta}N_{\rm mrg}/{\Delta}N_{\rm tot}$.
Similarly, we can obtain the fraction of BBHs end up with tidal disruption $f_{\rm td}$ 
and ionization due to the encounters with the background objects $f_{\rm ion}$.
These fraction of BBH are shown in Table~\ref{tab:model_frac}. 

\subsubsection{The Merging samples}
\label{subsubsec:mergingsamples}

We find that in galactic nulcei, typically $f_{\rm mrg}\sim3-40\%$ of the 
BBHs end up with merging events (See Table~\ref{tab:model_frac}). The highest fraction occur in clusters with MBHs 
around $\bh\sim10^6\msun$ and usually decrease for larger or smaller MBHs. 
When the inner orbit of a BBH in the numerical simulation is found dominated by GW radiation, 
the simulation ends and it is output from the {{\texttt{ GNC}}}. Usually the BBHs samples output from the 
simulations are with very different distributions of eccentricities and lifetimes~\footnote{
Across the paper, the ``lifetime'' of a BBH means the time duration from when the 
orbital evolution of BBH is dominated by the GW radiation to the final coalescence. The simulation of 
a BBH in the {{\texttt{ GNC}}} ends once the inner orbital evolution of the BBH is found dominated by the GW radiation,
i.e., when the GW decay timescale is much smaller than the timestep of the simulation.
For the details of timestep see Paper I.
}.
The results of MP1 are shown in Figure~\ref{fig:tgw_e_MP1} as an example. 

We found that these events have two unique features. The first one is that they have very significant 
eccentricities. As shown in Table~\ref{tab:model_frac}, about $40\%\sim70\%$ of them have extremely high eccentricities ($1-e_1\la 10^{-3}$). The other one is that some of the events can happen at positions very close 
to the MBH. For example, up to $10\%$ of all BBHs can merge at distance $r$ from the MBH, or 
 with SMA of the outer orbit $a$ smaller than $10^4r_{\rm SW}$. Up to $30\%$ of all merging events are with pericenter of the outer orbit $r_{p2}<10^4r_{\rm SW}$. Here $r_{\rm SW}$ is the Schwarzschild radius.
 
These merging BBHs can be either happens 
(1) After one or multiple encounters with the MBH (the fraction of which is 
denoted by $f_{\rm mrg}^{\rm eMBH}$). 
(2) after  Kozai Lidov oscillation (denoted by $f_{\rm mrg}^{\rm KL}$); 
(3) after one encounter with a background object (denoted by $f_{\rm mrg}^{\rm eBK}$);
Here $f_{\rm mrg}^{\rm eMBH}$, $f_{\rm mrg}^{\rm KL}$, and $f^{\rm eBK}_{\rm mrg}$ are all obtained 
by counting over the merging BBHs. 
\footnote{
For BBHs merge by (1) above, usually they merge right after the last close encounter of MBH, if there are multiple times of encountering. Note that the BBHs may not immediately merge during the process of (2) or (3), 
but merge later when they evolved isolatedly under gravitational orbital decay, 
due to the significantly excited eccentricity or small SMA resulting from their last dynamical 
impacts of (2) or (3). In these cases, the merging channel of a BBH is ascribed to the last dynamical 
impact it has ever experienced.}

A significant fraction (up to $f_{\rm mrg}^{\rm eMBH}\sim 25\%$) of all merging BBHs are merged after
one or multiple encounters with the MBH. These events happen
more likely around MBHs of $10^5\msun$. Before such encounter, the BBHs have usually experienced multiple 
encounters with background objects, epochs of KL oscillations and also 
encounters with the MBH. One example of such dynamical evolution is shown in
Figure~\ref{fig:track_fbhbmbh}. The eccentricity of the inner orbit of BBH is cumulatively excited 
from initial value of $e_1=0.236$ to $\sim 0.9$ before the final dramatic encounters with the MBH that lead to the 
merging event. The bottom panels of Figure~\ref{fig:track_fbhbmbh} show the trajectory (bottom left) 
and the evolution of both the inner and outer orbits (bottom right) right before the merging of the BBH.

The encounter of BBHs with MBH can excite the BBHs into extremely high eccentricities, i.e., 
up to $1-e_1\simeq 10^{-2}-10^{-6}$ (See Figure~\ref{fig:tgw_e_MP1}). 
For a few of them ($\sim1\%$), the eccentricity is so high that the BBH will plunge directly to each other 
with pericenter about a few times of horizon (which is a few to several tens of km), result in an rapidly decay 
GW transit event;
Mainly due to their high eccentricity, they can merge very quickly after the encounter with the MBH. 
Figure~\ref{fig:tgw_e_MP1} show that the lifetime of the majorities of these events are $10^3$s$\sim10^3$yr, 
the short ones of which merge within days right after the encounter. 

The relatively short lifetime of these events are understandable, as otherwise, in most cases, they will 
be modified again by a successive encounter with the MBH. Thus, the period
of the outer orbit of BBHs can set an upper limit of their GW lifetime. 
As a result, the lifetime of these events will be slightly longer if the mass of MBH 
is larger, e.g., $10^7\msun$, as shown in bottom left panel of Figure~\ref{fig:tgw_e_MP1}.

Another important feature is that 
these events merge very close to the central MBH. The merging event usually happens 
right after the encounter. For $\bh=10^5\msun$, the pericenter of the encounter can be about a few AU
(See the bottom right panel of Figure~\ref{fig:track_fbhbmbh}), and the merging position 
is at distance of about a few tens of AU from the black hole ($\sim 10^4r_{\rm SW}$ for $\bh=10^5\msun$). 
As these events happen very close to the MBH, the phases of GW will be significantly drifted due to the acceleration. 
We will discuss about them later in Section~\ref{sec:relativity_BBHs}.

{
Note that the merging of a BBH is usually a result of multiple encounters with the MBH.
As the outer orbit of a BBH is dynamically evolving under the two-body and resonant
relaxations, a BBH will experience multiple encounters (which can be up to 
hundreds and thousands) with the MBH when it moves from $D=r_p/r_{\rm td}\sim3$ into $D\sim1$
before being tidally disrupted, where $r_{td}$ is the tidal radius of the BBH. 
For example, the BBH illustrated in Figure~\ref{fig:track_fbhbmbh} is merged after in 
total of $45$ times of encounters with MBH. The probability of a BBH with inital eccentricity $e_1<0.9$
merging after encountering with MBH with 
$D<3$ is usually $0.01\%-0.1\%$ per encounter. In most cases the BBH will remain integrity after that although 
the inner orbital eccentricity is changed. Thus, the probability of merging is dramatically increased 
due to the excited inner eccentricity of the BBH and the increased number of encounters. 
}

\begin{figure*}
\center
\includegraphics[scale=2.4]{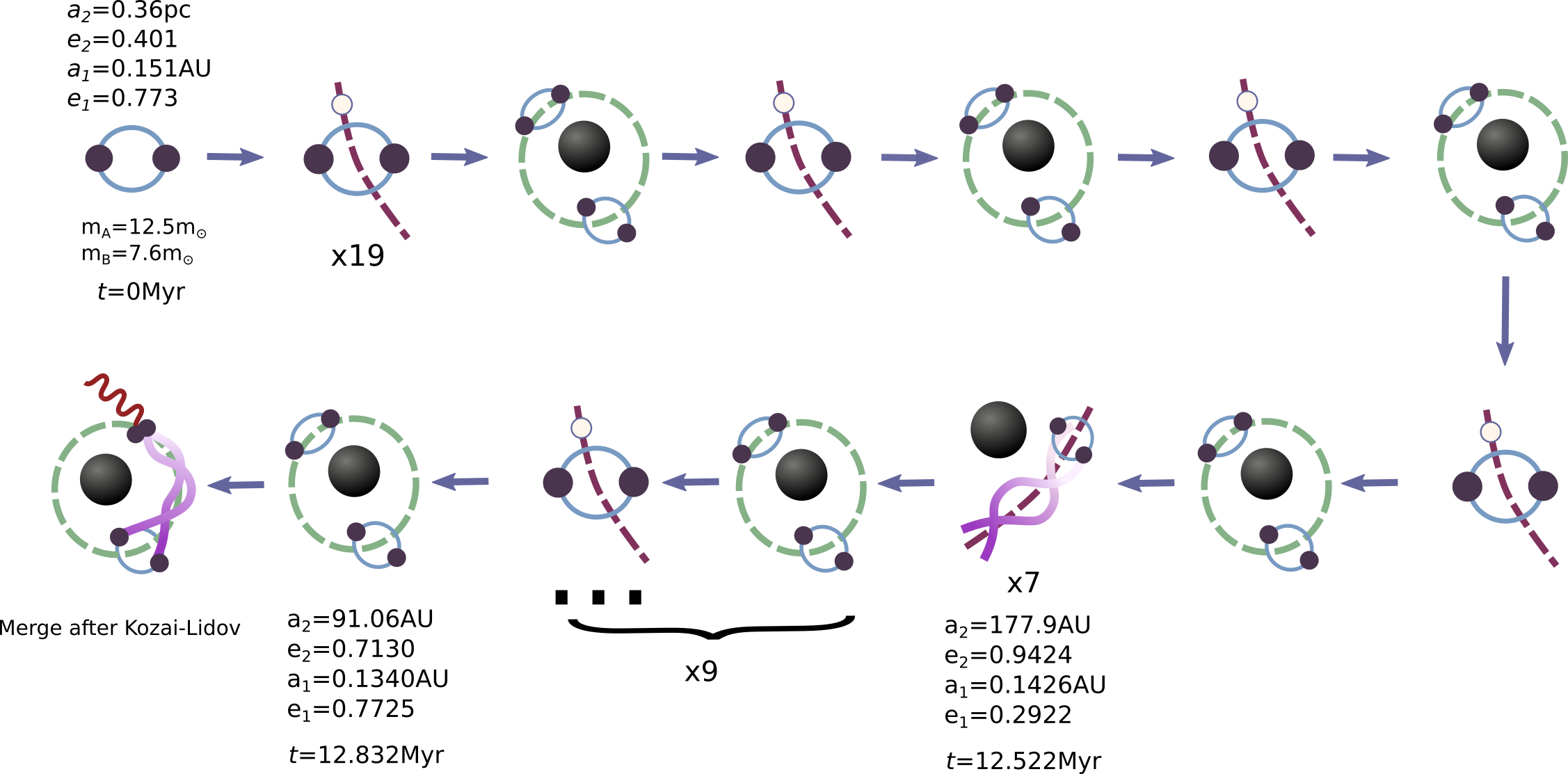}
\includegraphics[scale=0.42]{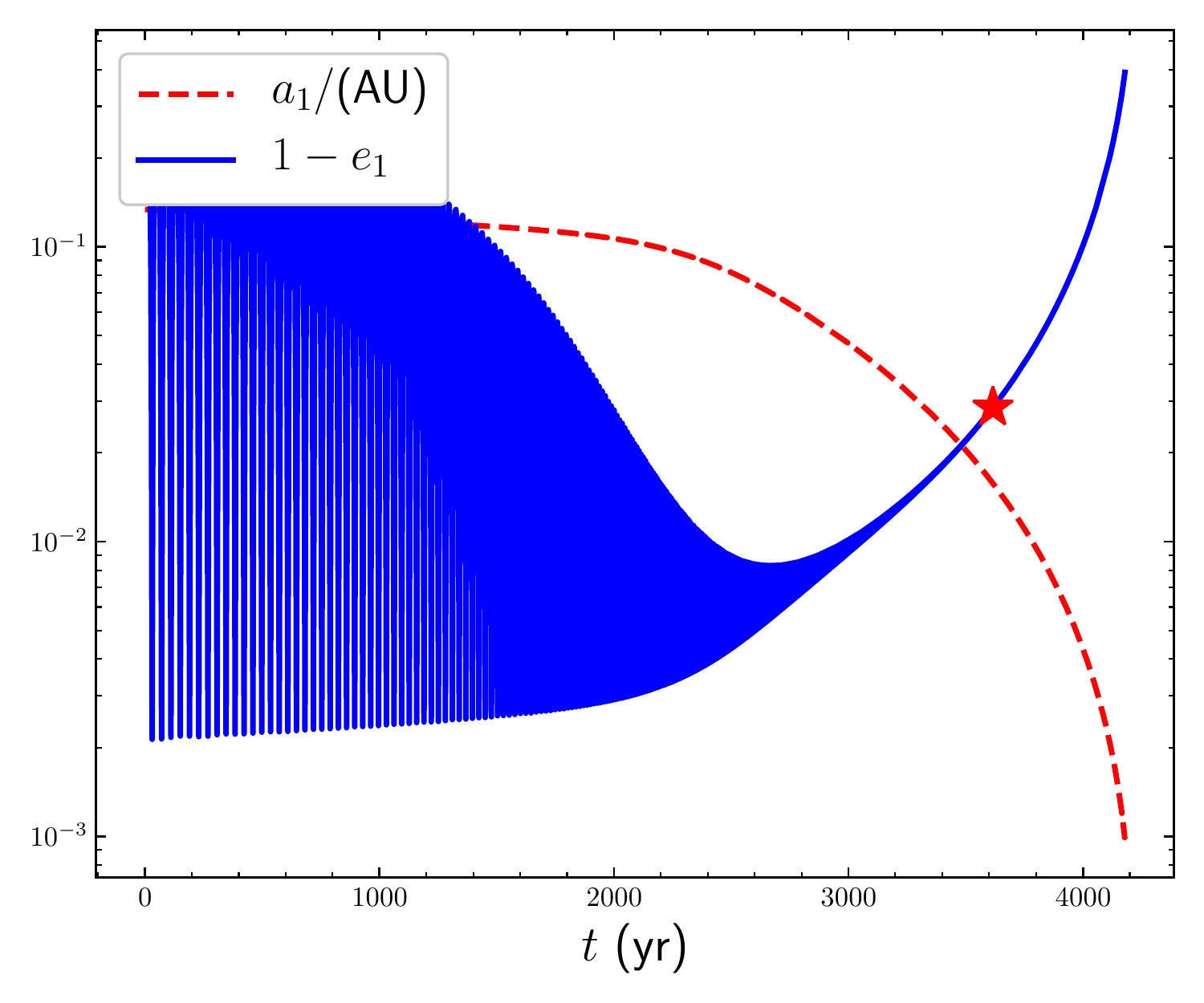}
\caption{The illustrations on the left show one example of the 
dynamical evolution history of a BBH that merge due to KL oscillation. 
The example is from model MP1 of simulations runs given $\bh=10^5\msun$. 
The illustration of each dynamical processes are similar to Figure~\ref{fig:track_fbhbmbh}.
The orbital evolution of the inner orbit of BBH during the final KL oscillation are shown in the panel on the right.
The red star in the eccentricity curve marks the 
position where the {{\texttt{ GNC}}} find that the GW orbital decay dominates the evolution of the inner orbit of the BBH.}
\label{fig:dy_evl_kl}
\end{figure*}

Another important merging channel of BBHs is the KL oscillation, of which the fraction 
to all the merging events can be up to $\sim70\%$, and decrease to $\sim6-16\%$ with increasing mass of 
MBH. Similarly, KL and other dynamical processes have been cumulatively modified the BBHs before the 
last KL oscillation that leads to the final merging event. One example of such evolution is illustrated in Figure~\ref{fig:dy_evl_kl}. The panel on the right of Figure~\ref{fig:dy_evl_kl}
show the evolution of the inner orbit of BBHs before the final merging event. The {{\texttt{ GNC}}}
output the eccentricity of this event with $e_1\sim 0.97$ when GW dominates the evolution of the inner orbit. 
As shown in Figure~\ref{fig:tgw_e_MP1},  the output eccentricity of BBHs due to KL can be up to
$1-e_1\sim10^{-7}$, comparable to those due to BBH-MBH encounters.
The lifetime of these events are about a few hours up to $10^{5}$ yr, slightly depends on 
the mass of the MBH.

The encounters with the background objects is the third merging channel of BBHs, 
the fraction of which varies between $7.0\%\sim90\%$ of all merging events\footnote{Some of the BBHs will harden and become 
unbound to the MBH due to the three-body encounter. 
However, as the escaping velocity is usually small ($\la$ tens of $\kms$), them are still considered 
bound in the cluster or the galactic nucleus. }. The rates 
are higher if increasing the mass of the MBH. The resulting 
eccentricities are relatively smaller compared to the first two channels mentioned above, 
i.e., $1-e_1=10^{-4}-1$ (See Figure~\ref{fig:tgw_e_MP1}). The lifetime of these events can 
be up to $10^4$yrs around $10^5\msun$ MBH or up to $10^7$ yrs around $10^8\msun$ MBH. 
In most cases, such a long time is limited by the dynamical timesteps $\delta t$ of {{\texttt{ GNC}}}
as we require that the merging timescale of these BBH samples satisfy $t_{\rm GW}<0.1\delta t$.
Here $\delta t$ is set according to the timescales of two body relaxation, resonant relaxation, or the 
collision rate between BBHs and the background objects. For more details, see Paper I.

The overall merging rates in the local universe can be estimated according to 
Equation~\ref{eq:Rate_tot}, and the results can be seen in Table~\ref{tab:model_frac}. 
We find that the rate mainly depend on the assumed background masses of the stars. 
For models assuming $m_\star=1$ (models MP1-3 and MP5), the estimated rates are 
$5-200$Gpc$^{-3}$yr$^{-1}$. If the background mass is $m_\star=10\msun$ (model MP4), 
the rate reduce to $0.2-1.7$Gpc$^{-3}$yr$^{-1}$. When the background mass is mixing 
stars and stellar black holes (in model MP6), the rate is between the above two, 
i.e., $7-71$Gpc$^{-3}$yr$^{-1}$. Thus, all models except MP4 are consistent with the 
rate given by aLIGO/Virgo detection, which is $\sim 23.9$Gpc$^{-3}$yr$^{-1}$\citep{2021ApJ...913L...7A}. 
As the number and masses 
should be dominated by main-sequence stars over the entire cluster except the very inner regions
~\citep{Alexander05}, MP4 should be unfavored by the observations. 

\subsubsection{The tidally disrupted or ionized samples}
The BBHs may also end up with tidal disruption by close encounters with the
MBH or be ionized due to encounters with the background objects. 
$f_{\rm td}$ can be very significant near small mass MBH 
($\sim 60\%$) and decreased to $\sim7\%$ for MBH with $10^8\msun$.
This is because in small mass MBH, the two body and resonant 
relaxation dynamical timescale is 
much shorter, making them more easily move into the loss cone region.  
The fraction of ionization of BBHs is low, i.e., $f_{\rm ion}\sim 1-7\%$, 
if the masses of the background stars are small (e.g., MP1-3 and MP5). 
However, $f_{\rm ion}\sim 10-20 \%$ if the background stars' masses
is $m_\star=10\msun$ (for MP4), and $f_{\rm ion}\sim 5-11\%$ if the background objects mix with single 
stars and black holes. These results suggest that the tidal 
forces of the MBH, rather than the encounters with the background stars, is 
more effective in destroying the binarities of black holes in the vicinity of MBH.

\section{Observing BBHs in current and future GW observatories}
\label{sec:OB_LISA_LIGO}
\begin{figure*}
\center
\includegraphics[scale=0.65]{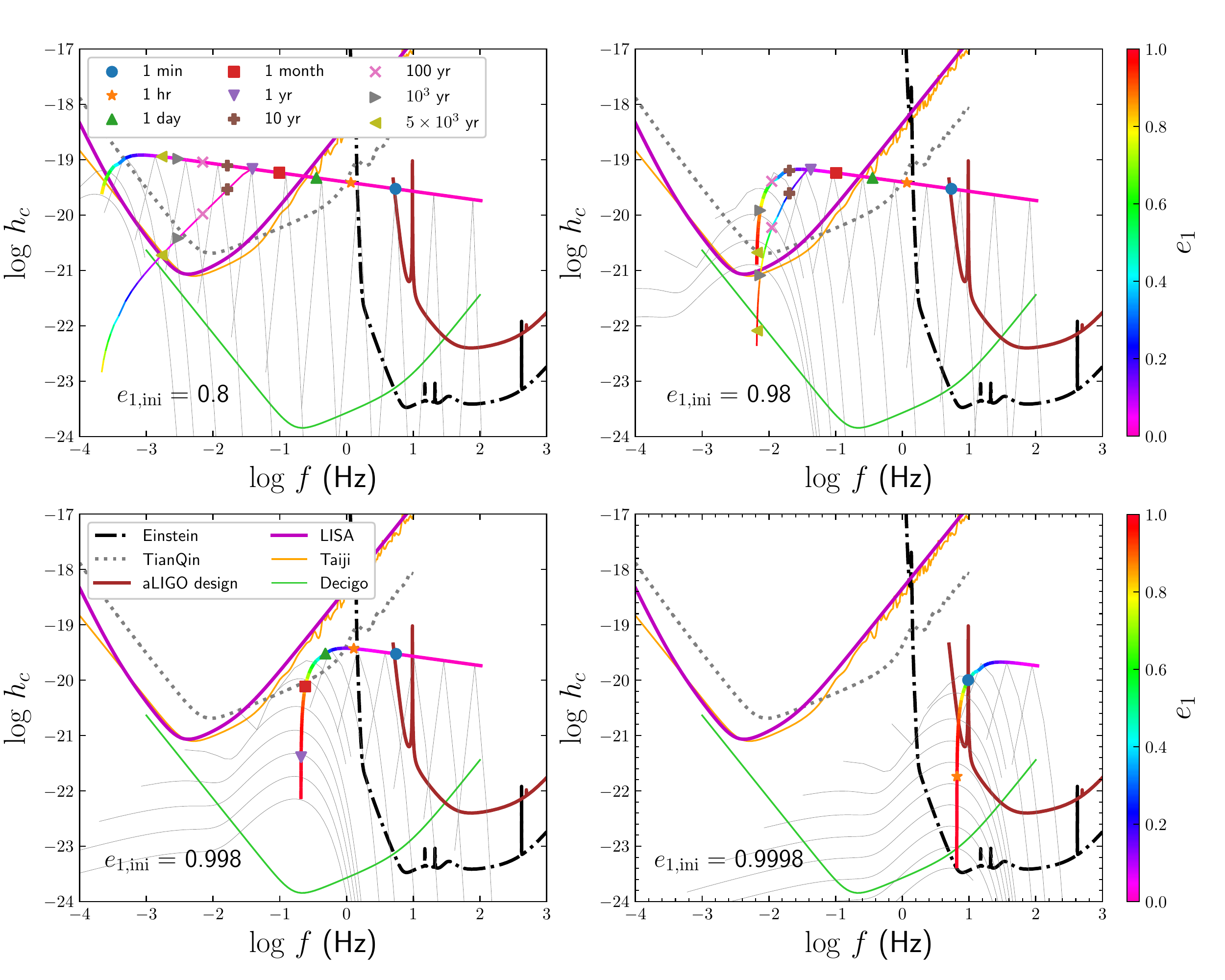}
\caption{The evolution of the strain amplitude $h_c$ for BBHs with different eccentricities. 
BBHs start with orbital period $P_{1,\rm ini}=1$day and with different eccentricity ($e_{1,\rm ini}$ in each panel). For illustration purpose, all BBHs are with $m_A=m_B=20\msun$ and $z=0.01$. The noise-level 
of different GW observatories are shown in lines with different color and styles, see the legends in the lower right panel. The thin gray lines show the GW spectrum at different
harmonics. The thick (or thin)
solid lines are the evolution of the peak of the GW radiation (or assuming that the BBHs have no frequency evolution in a $t_M=5$ yr's mission, see texts in Section~\ref{sec:OB_LISA_LIGO} for more detail), 
of which the color map show the eccentricity ($e_1$) of the BBHs during the evolution. 
}
\label{fig:harmonic_full_ob}
\end{figure*}
In this section, we investigate how the merging BBHs obtained by {{\texttt{ GNC}}} described 
in Section~\ref{sec:evl_in_cluster} appear in the current (aLIGO/Virgo O2) and future GW observatories, i.e., aLIGO (design),
Einstein, { DECIGO}, TianQin, LISA and Taiji observatories. 
We mainly focus on the evolution and harmonics of the BBHs appeared in the observations (Section~\ref{subsec:evl_harmonic}), 
and we investigate the numbers and SNR of the inspiralling and merging of BBHs (Section~\ref{subsec:mc_inspiral} and Section~\ref{subsec:snr_BBHs}).

\subsection{The evolution of BBHs and the harmonics of GW}
\label{subsec:evl_harmonic}
As shown in Section~\ref{subsec:evl_merging_BBHs} and Figure~\ref{fig:tgw_e_MP1},
BBHs form in galactic nuclei around MBH usually have very significant 
eccentricities, thus their GW spectrum will spread to high orders of harmonics.
For each of the BBH merging events from our simulation, we can get their 
 orbital evolution under GW orbital decay and also the characteristic strain of GW at the 
 $n$th harmonic before coalescence, which is given by~\citep{2004PhRvD..70l2002B}
(See also~\citet{2021arXiv210316030R} for a more efficient way of estimation.)
\be
h_{c,n}(f)=\frac{1}{\pi D}\sqrt{\frac{2G}{c^3}\frac{dE_n}{df_r}}
\label{eq:hcn}
\ee
Where $D$ is the comoving distance of the BBH. 
The detail forms of $dE_n/df_r$ can be found in the 
literature~\citep[e.g.,][]{2015PhRvD..92f3010H,2017MNRAS.470.1738C, Peters63}. 
We summarize them in the Appendix~\ref{apx:GW_harmonic}. 

The SNR is then given by~\citep{Oleary09}
\be
\rho^2=\left(\frac{\rm S}{N}\right)^2=\sum^{n_{\rm max}}_{n=1}\int^{f_{\rm max}}_{f_{\rm min}}
\frac{h_{c,n}^2(f)}{fS(f)}\frac{df}{f}
\label{eq:snr_estimate}
\ee
where $f=f_0(1+z)$ and $f_0$ is the frequency of the GW in the rest frame and in the observer's frame, 
respectively; $S(f)$ is the strain spectral sensitivity of the observatory. 
For aLIGO/Virgo(O2) and aLIGO(design) sensitivity, it is given by 
\citet{Barsotti12} and \citet{Barsotti18}, respectively. For LISA $S(f)$ is given by~\citet{2019CQGra..36j5011R}, 
Einstein by~\citet{2011CQGra..28i4013H}, { DECIGO by~\citet{2011PhRvD..83d4011Y}}, 
TaiJi by~\citet{2020IJMPA..3550075R} and TianQin 
 by~\citet{2016CQGra..33c5010L, 2019PhRvD.100d3003W}.

The integrations are performed between the minimum and maximum frequency of the 
BBHs during the mission, and they depend on the order of harmonic $n$. 
For a given BBH, if the minimum and maximum of GW frequency at 
$n$-th harmonic in a five years mission is given by $f_{{\rm min}, n}$ and $f_{{\rm max}, n}$, 
respectively, the minimum and maximum frequency of the observational band are given 
by $f_{\rm min, ob}$ and $f_{\rm max, ob}$, respectively, 
then the lower and upper limit for the integration is given by
$f_{\rm min}={\rm max}(f_{{\rm min}, n}, f_{\rm min, ob})$,
$f_{\rm max}={\rm min}(f_{{\rm max}, n}, f_{\rm max, ob})$, respectively.

The duration and evolution of the GW of a BBH observed in an observatory depend largely on its 
eccentricity. Figure~\ref{fig:harmonic_full_ob} illustrates some of the examples of BBH's $f-h_c$ evolution. 
For a BBH starts with $P_{1,\rm ini}=1$day, 
and $e_{1,\rm ini}=0.8$, the strains enter into LISA band when its eccentricity 
decreases to $e_1\sim0.4$, and then stay in LISA band for thousands of years before merging. 
Most of these events will be considered as background sources in LISA band as they evolved too slowly unless they have significant SNR. For these BBHs with almost no frequency evolution,
Equation~\ref{eq:snr_estimate} become $\rho^2\simeq \sum h^2_{c,n}
(\dot{f} t_M/f)/(fS(f))$, where $t_M$ is the mission time. Thus, when $t_M<f/\dot f$, we can 
approximately consider that the BBHs have no evolution in frequency, and the quantity $h_{c,n}
\sqrt{\dot{f} t_M/f}=h(f)ft_M$ (thin solid line with colormap in Figure~\ref{fig:harmonic_full_ob}) 
provides an effective characteristic amplitude that can be used 
in estimating SNR, where $h(f)$ is the amplitude of the GW~\citep{2000PhRvD..62l4021F, 2018MNRAS.481.4775D,2019CQGra..36j5011R,2005ApJ...623...23S}. 

If the eccentricity is higher, i.e., $e_{1,\rm ini}=0.98$, the strain amplitude rise above the noise level of 
LISA with slightly increased GW frequency when $e_{1}\sim 0.8$,
and then moves towards Einstein and aLIGO band after evolving about a thousand years. 
If $e_{1,\rm ini}=0.998$, the evolution become much faster and 
the peak arises directly into the band of TianQin, without passing through LISA band, and then 
move into Einstein and aLIGO band within one month. If a BBH has an even higher eccentricity of
$e_{1,\rm ini}=0.9998$ ($1-e_{1,\rm ini}\sim10^{-4}$), its characteristic strain can directly rise above 
and into Einstein and aLIGO band. These results suggest that for very highly eccentric BBHs, 
they can only be seen in detectors with entering frequencies larger than $0.1\sim 1$ Hz 
(e.g., aLIGO, Einstein or DECIGO detectors).

\subsection{Generating samples of inspiralling/merging BBHs}
\label{subsec:mc_inspiral}
\begin{figure}
\center
\includegraphics[scale=0.7]{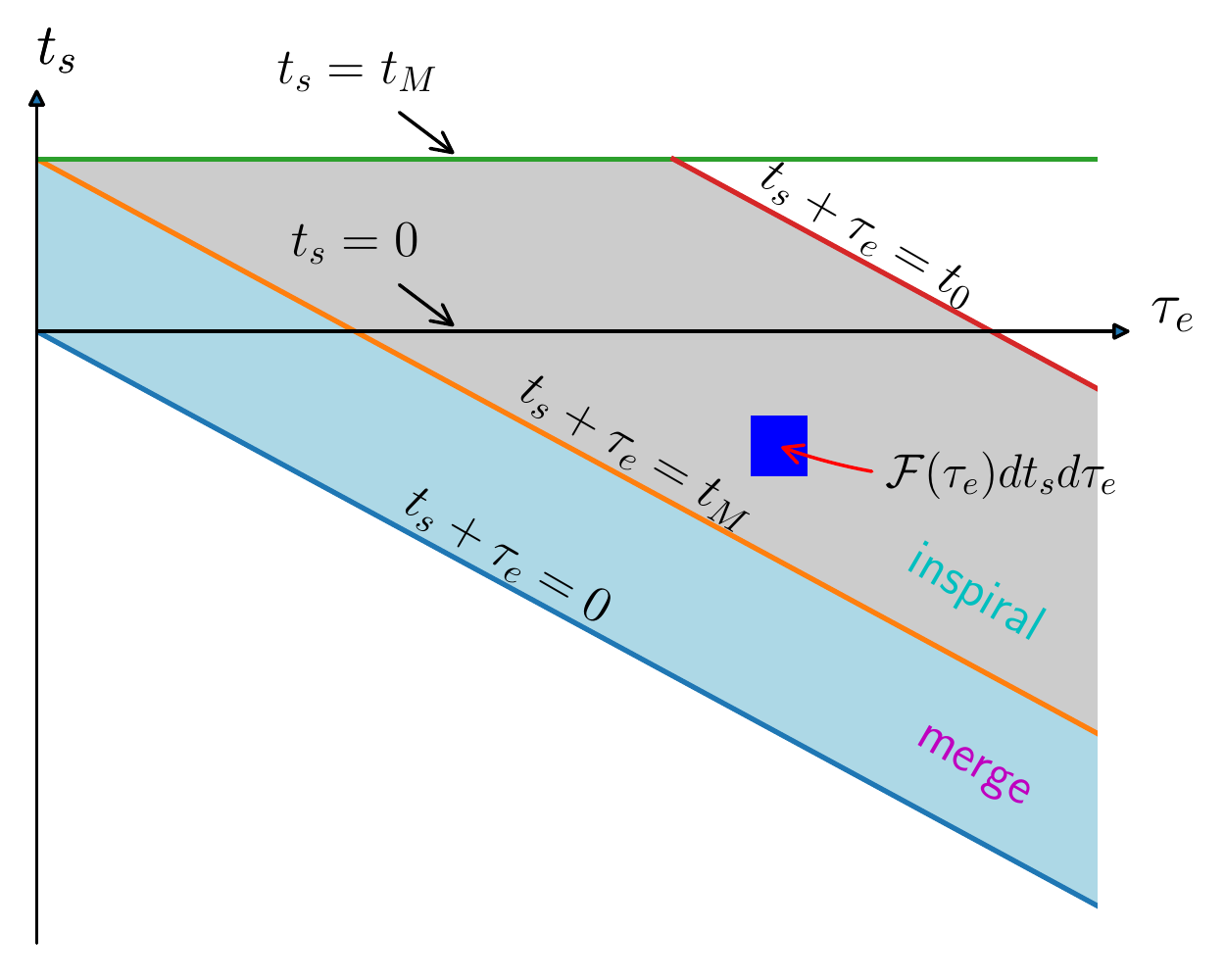}
\caption{Figure show the timeline of the generation and the merging of samples. $t_s$ is the 
time where sample forms and $\tau_e$ is the lifetime of each sample observed on earth.
The observation starts at $t_s=0$ and ends at $t_s=t_M=5$ yr. 
The samples with $t_s<-\tau_e$ are already coalesced before the start of the mission, and 
those form between $-\tau_e<t_s<t_M-\tau_e$ filled by color light blue are samples 
coalescing during the mission. 
The regions filled by light grey are samples insprialling with time of coalescence 
$t_{\rm CO}$ less than a given time of interest $t_0$ during the mission. 
$\mathcal{F}(\tau_e)$ is the formation rate of samples with lifetime $\tau_e$ given by 
Equation~\ref{eq:ftau_e}.}
\label{fig:timeline}
\end{figure}

\begin{figure*}
\center
\includegraphics[scale=1.0]{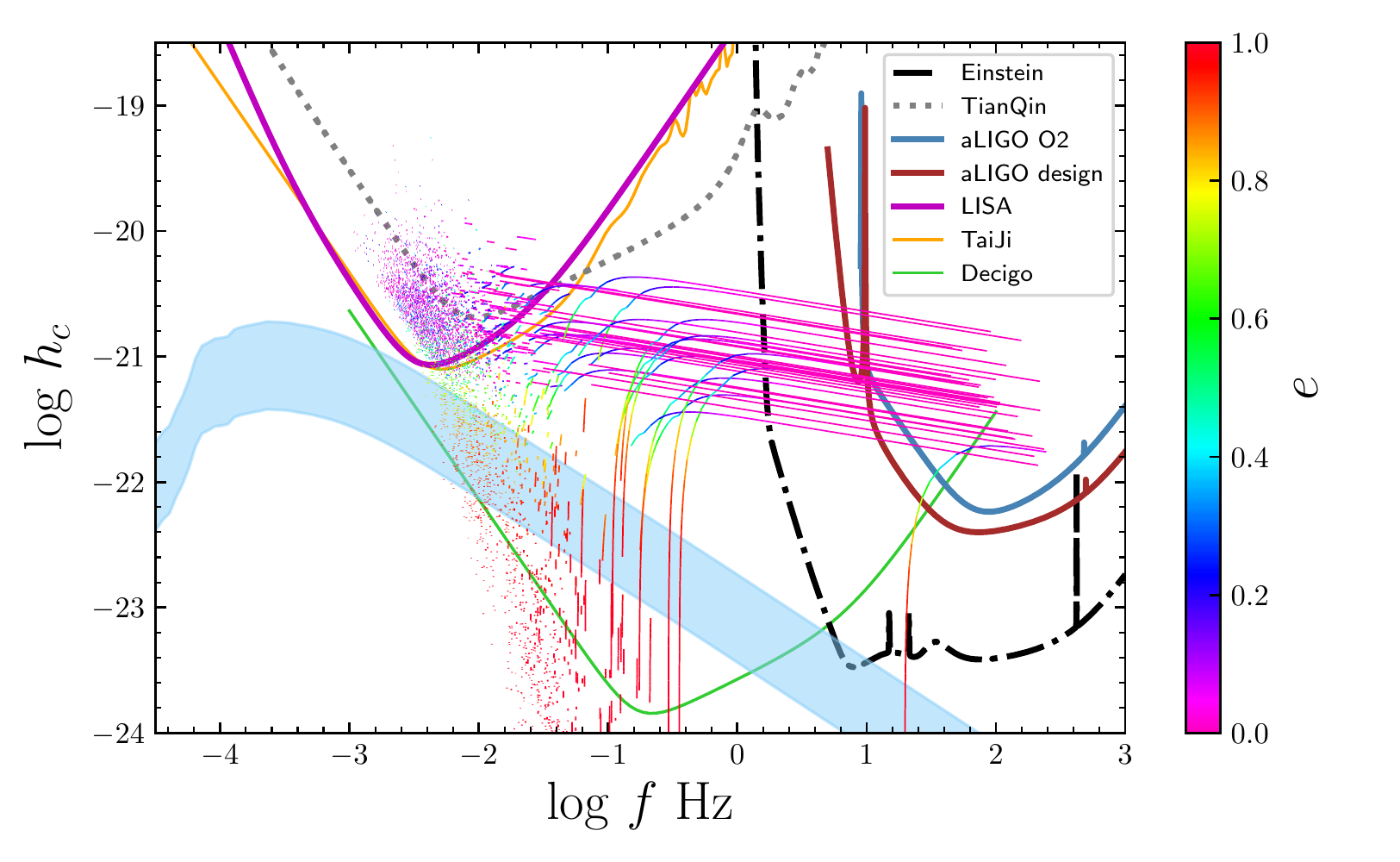}
\caption{The $5$yr evolution of the strain amplitude of inspiralling BBHs from a MC
realization described in Section~\ref{subsec:mc_inspiral}, by using samples of MP1 (in total of $5000$ samples are presented as examples). For clarity of the figure, here we only show samples with $0.001<z<0.5$, and 
with time to coalescence at the start of the observation less than $10^3$ yr. 
The solid lines with color maps are the evolution of the peak of the GW radiation. 
The color map in the solid line show the eccentricity $e_1$ of the BBHs during the evolution. 
The light-blue shaded region show the characteristic strain noise amplitude of the GW background 
from BBHs merging around MBHs. For more details of the GW background, see Section~\ref{sec:GWBK}. }
\label{fig:mc_sample_all}
\end{figure*}

\begin{figure*}
\center
\includegraphics[scale=0.7]{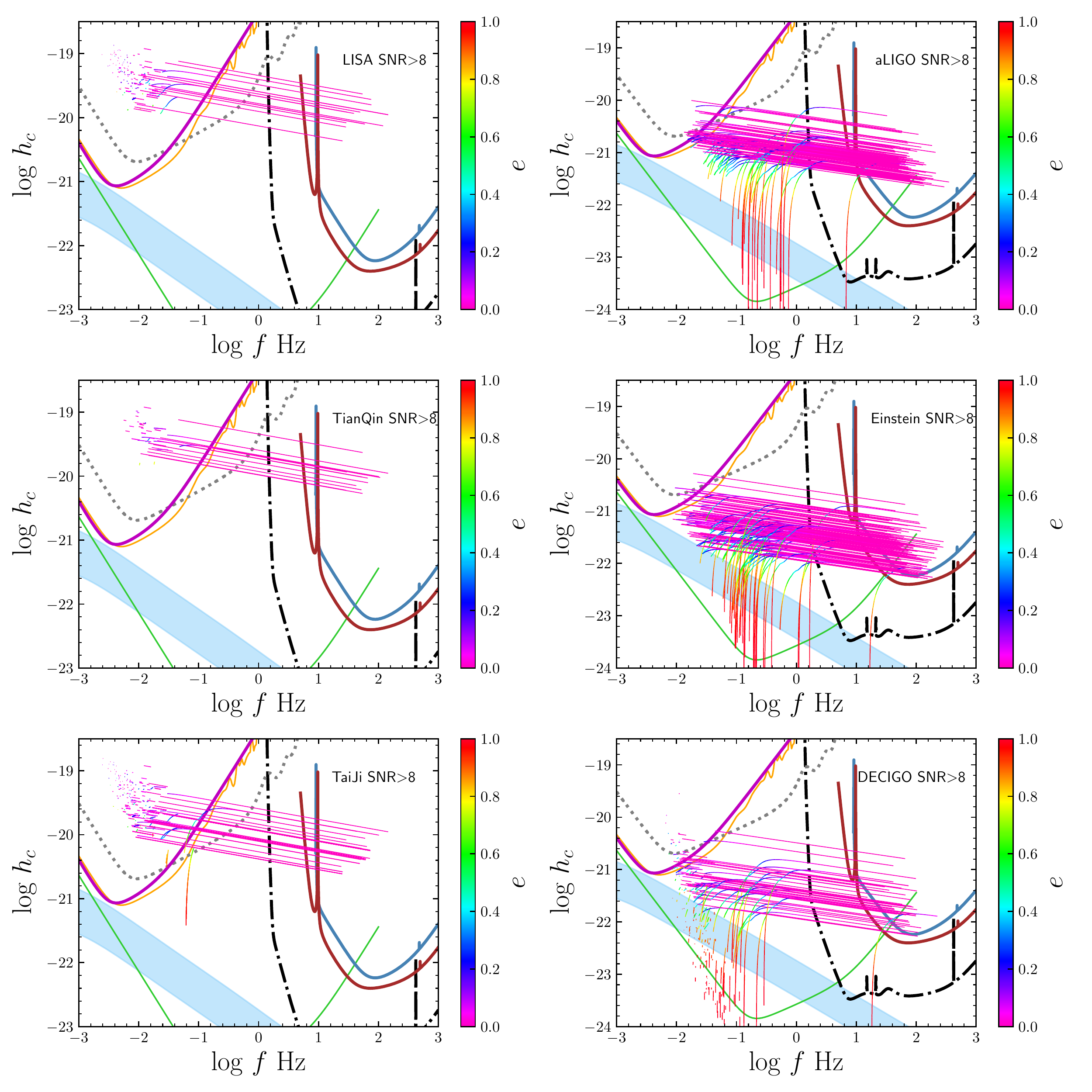}
\caption{Examples of the evolution of BBHs with SNR$>8$ in { LISA, aLIGO (design),
TianQin, Einstein, TaiJi and DECIGO} during a continuous $5$ yr mission.
The solid lines with color maps are the evolution of the peak of the GW radiation. 
The color map in the solid line show the eccentricity $e_1$ of the BBHs during the evolution. 
The samples are limited with redshifts of $0.001<z<2$. 
}
\label{fig:mc_sample_ob}
\end{figure*}

Section~\ref{subsec:merging_rate} and Table~\ref{tab:model_frac} provide the merging rates of BBHs in the local 
universe. These numbers are related to the final stage of the merging BBHs, thus usually useful for aLIGO/Virgo detection.
However, LISA or other low frequency detectors can probe the inspiralling phase of 
BBHs. The number of inspiralling BBHs detected in the different observatories should 
depend on the mission time ($5$ yr in this work), 
the evolution trajectories of $f-h_c$ of the individual BBHs, 
the detection sensitivity of observatories, and etc. 

In order to investigate the inspiralling samples that can be observed in individual or 
mulitiple observatories, first we need to estimate the number of inspiralling/merging BBHs given
some distribution of their lifetime (defined in Section~\ref{subsubsec:mergingsamples})
and then generate a group of Monte-Carlo samples for SNR estimation. The details are in the following sections. 

\subsubsection{Estimating the total number of inspiralling/merging BBHs}
\label{subsubsec:num_insp_merge}
Suppose that we have a group of samples resulting from a simulation by {{\texttt{ GNC}}} given MBH mass of $\bh$, 
and denote the local lifetime of samples from the simulation by $\tau$ (The time span of a BBH 
from when GW dominate its inner orbital evolution to its final coalescence, see also Figure~\ref{fig:tgw_e_MP1}), then 
the formation rate of samples with $\tau\rightarrow\tau+{\rm d}\tau$ is given by 
$\mathbf{R}({\bh},\tau){\rm d}\tau=\mathbf{R}(\bh)f(\tau|\bh){\rm d}\tau$, 
where $f(\tau|\bh)$ is the probability distribution function (pdf) of $\tau$ from the simulation given MBH mass of 
$\bh$ and $\mathbf{R}(\bh)$ can be estimated by Equation~\ref{eq:RS}. 
The formation rate of BBHs observed per unit time of earth with per unit 
lifetime $\tau_e=(1+z)\tau$ observed on earth, is given by integrating over all MBHs and all redshift:
\be\ba
\mathcal{F}(\tau_e)= \int_0^\infty \frac{n(\bh, z)}{(1+z)^2} 
 R\left(\bh,\frac{\tau_e}{1+z}\right) d\bh \int_0^\infty\frac{dV_c}{dz} dz
 \label{eq:ftau_e}
\ea\ee
where $n(\bh, z)$ is the black hole number density per unit comoving volume given by~\citet{2015ApJ...810...74A}, and 
$V_c$ is the comoving volume. Then the number of BBH samples formed measured at earth time $t_s\rightarrow t_s+dt_s$ 
with lifetime between $\tau_e\rightarrow\tau_e+d\tau_e$ is given by $\mathcal{F}(\tau_e)dt_sd\tau_e$
\footnote{We have already assumed that the maximum time of lifetime of each BBHs is much smaller than 
the cosmic time interval that corresponding to $dz$. 
Also, we have assumed that the formation rate $\mathcal{F}(\tau_e)$ for a given $\tau_e$ 
is a constant as function of time on earth during the mission. 
}. Note that here $t_s$ is the formation time of BBHs observed on earth (the moment when the evolution of 
inner orbit of BBH is dominated by GW radiation). Here we consider redshift between $0.001<z<2$.

The formation and coalescence timeline of samples with different $\tau_e$ can then be illustrated
in Figure~\ref{fig:timeline}. Suppose that $t_s=0$ is the time when the observation starts, then
samples form before $t_s\le -\tau_e$ coalesce at time $t_s\le0$, right before the start of the mission. 
Those form between $-\tau_e<t_s<t_M-\tau_e$ will coalesce during the mission, where $t_M$ is the mission time. 
If the time to coalescence of a BBH measured at the start of the mission is denoted by $t_{\rm CO}$, 
then the number of samples merging, or inspiralling with $t_{\rm CO}$ less than a time of 
interest $t_0$, i.e., $t_{\rm CO}<t_{0}$ ($t_0=1000$yr, see texts later) can be obtained by :
\be\ba
N_{\rm tot}&(t_{\rm co}<t_{0}| t_M)=N_{\rm merge}(t_M)+N_{\rm insp} (t_{\rm co}<t_{0}| t_M)
\label{eq:Ninsp}
\ea\ee
where $N_{\rm merge}$ is the number of merging samples during the mission,
and $N_{\rm insp}$ is the number of insprialling samples with time to coalescence  
$t_{\rm CO}<t_0$ at the start of the mission. According to Figure~\ref{fig:timeline}, 
they can be estimated by  
\be\ba
N_{\rm merge}&=t_M \int^{\infty}_{0} \mathcal{F}(\tau_e) d\tau_e\\
N_{\rm insp} &=\int^{t_0-t_M}_0 \tau_e\mathcal{F}(\tau_e) d\tau_e
+t_0\int^\infty_{t_0-t_M} \mathcal{F}(\tau_e) d\tau_e
\ea\ee

From Figure~\ref{fig:harmonic_full_ob}, we can see that most BBHs appeared in LISA band usually 
have a lifetime of $\la 10^3$yr. Thus, by setting $t_0=1000$yr is sufficient in covering most BBHs that 
can be detected with high SNR in ground and space-based observatories. 
We found that for models in Table~\ref{tab:model}, in a $5$ yr mission, 
the typical number of $N_{\rm tot}(t_{\rm co}<10^3{\rm~yr})\sim 10^6$. 

The total number of inspiralling samples at any given moment are given by $t_M\rightarrow0$
and $t_0\rightarrow\infty$, which can be further reduced to
\be
N_{\rm tot}=\int^\infty_0 R(\bh, \tau)\tau d\tau \int^\infty_0 dz \int^\infty_0 d\bh n(\bh, z)\frac{dV_c}{dz}
\ee
We find that the typical number for all models is $N_{\rm tot}\sim 10^9$. Most of these sources should 
be very slowly evolving and contributing to a background of GW. The amplitude of the GW backgrounds
contributed from BBHs merging around MBH will be discussed later in Section~\ref{sec:GWBK}.

\subsubsection{Generating the inspiralling/merging samples by Monte-Carlo method}
We use a Monte-Carlo (MC) method to generate the mock inspiralling/merging samples during a given time of mission
$t_M$ and with time to coalescence less than $t_0$. Statistical analysis of these samples can 
then provide predictions of the inspiralling/merging samples for different models. Each of them is generated following the steps below:
\begin{enumerate}
\item Draw a pair of $\bh, z$ from a two-dimensional distribution given by
\be
P(\bh, z)\propto n(\bh, z)\frac{\mathbf{R}(\bh)}{1+z}\frac{dV_c}{dz}
\ee

\item For a $\bh$ from above, get the samples from the simulation runs of {{\texttt{ GNC}}} given MBH mass 
closest to $\bh$ (one of the $\bh=10^5\msun, 10^6\msun, 10^7\msun$ and $10^8\msun$). Randomly select one of the samples, get its corresponding lifetime observed on earth, i.e., $\tau_{e}=\tau (1+z)$. 
According to Figure~\ref{fig:timeline}, suppose that the mission last for $t_M=5$ yr, and 
we are only interested for samples with time of coalescence less than $t_0=1000$yr
at the start of observation, then the probability of accepting this sample 
with lifetime $\tau_{e}$ is $\propto {\rm min}[1,(t_M+\tau_e)/t_0]$. If the sample is 
rejected, then go back to step (1).

\item The formation time of this sample observed on earth ($t_{s}$) is uniformly distributed between 
$(-\tau_{e}, t_0-\tau_{e} )$ 
if $\tau_{e}>t_0-t_M$, or between $(-\tau_{e},t_M)$ if $\tau_{e}<t_0-t_M$.
%\item Given the $t_s$, the evolution time before the start of the observation is then $t_{\rm evl}=-t_{s,i}/(1+z_i)$ 
%if $t_{s,i}<0$ at the local universe. For samples with $t_{s,i}>0$, they are newly borned binaries.
\item At the start of the mission, if $t_{s}<0$,  the time of coalescence of each sample 
is given by $t_{\rm co}=(t_{s}+\tau_{e})$; If $t_{s}>0$, they are newly formed samples and their time of colascences is given by $t_{\rm co}=\tau_e$.
\end{enumerate}
Such process repeats until a sufficiently large number of Monte-Carlo samples are realized.
The orbital evolution of these samples due to GW orbital decay are then calculated and 
the frequency and the strain amplitude of GW are estimated according to Equation~\ref{eq:hcn}

Figure~\ref{fig:mc_sample_all} show the $5$ yr evolution of $5000$ examples resulting 
from above MC simulation for model MP1.
We can see that in LISA/TaiJi band, many of the BBHs inspiral at frequency 
$1\sim10$mHz do not show significant evolution during the mission time. 
Only samples above $10\sim100$mHz show 
some evolution towards high frequencies. The eccentricities of a
number of BBHs are so high that, the $h_c$ rise up at frequency $\sim0.1$Hz and 
directly evolved into Einstein and aLIGO/Virgo band without passing through any of the space-based observatories such as LISA/TianQin/TaiJi. We expect that such a rising phase of $h_c$ 
can only be probed by { DECIGO}~\citep[see also][]{2017ApJ...842L...2C}, 
that are sensitive to frequencies between $0.1-10$Hz~\citep{2019arXiv190811375A,2020MNRAS.496..182L}.

\subsection{The SNR of BBHs}
\label{subsec:snr_BBHs}
             
\begin{figure*}
\center
\includegraphics[scale=0.65]{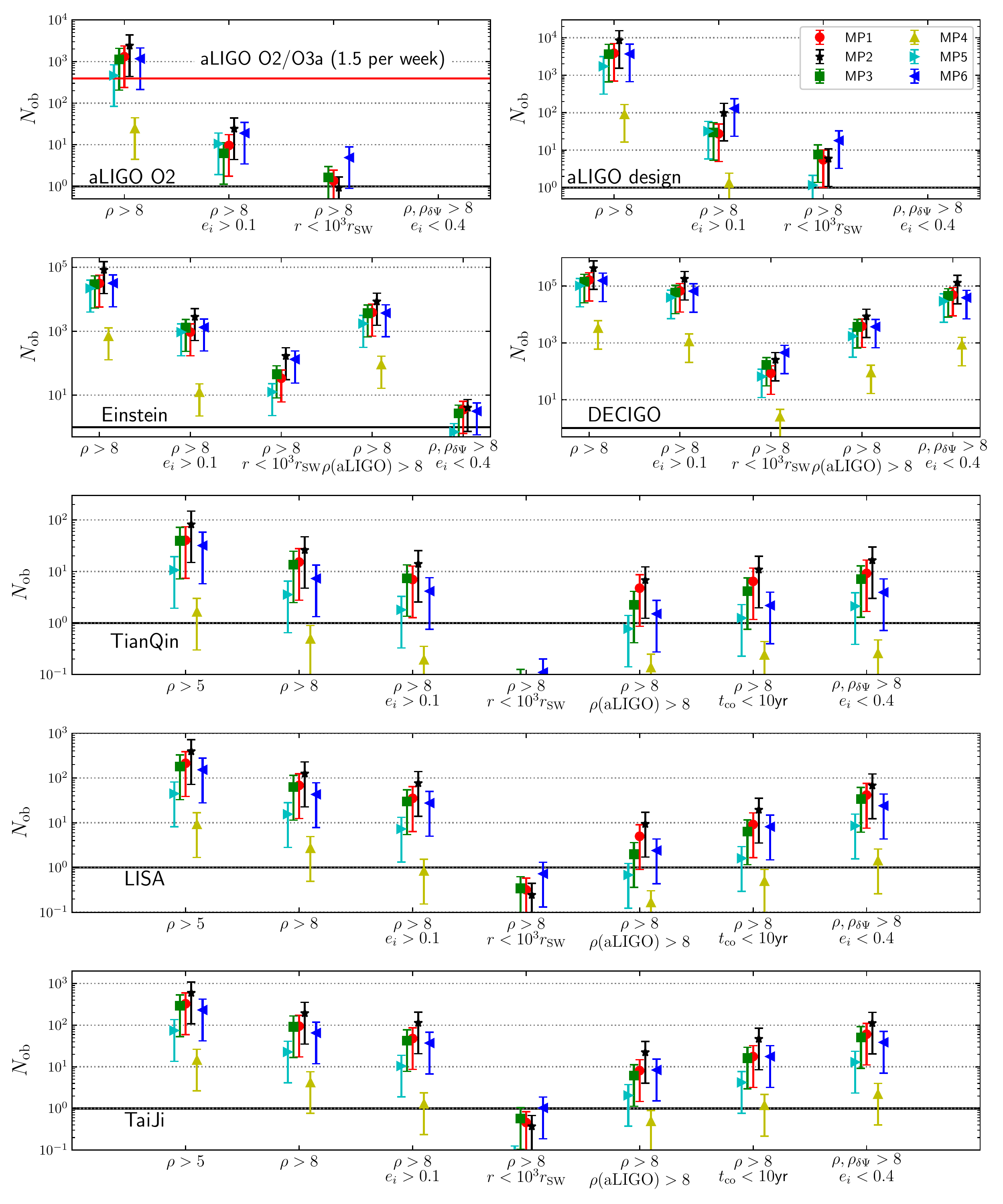}
\caption{Predicted numbers of inspiralling/merging BBHs in different observatories given SNR $\rho>8$ (or $\rho>5$)
and also some other conditions. 
For example, with entering eccentricity $e_i>0.1$, 
or inspiralling/merging at distance of $r<10^3r_{\rm SW}$ ($r_{\rm SW}$ is the Schwarzchild radius) to the MBH, 
or with the SNR of the Doppler drift of phase $\rho_{\delta \Psi}>8$ (See Equation~\ref{eq:snr_phase_drift}) 
and with $e_i<0.4$, or with time to coalescence at the start of mission less than ten years ($t_{\rm co}<10$yr). 
We also predict number of BBHs that can be observed in multiband observations. For example, 
we show the number of samples with $\rho>8$ in Einstein, LISA, TianQin and TaiJi and 
also with SNR$>8$ in aLIGO(design) band ($\rho(\rm aLIGO)>8$). Symbols 
for each model show the mean value of the prediction and the errorbar 
is from the uncertainty of the event rates. }
\label{fig:Nob}
\end{figure*}

Here we mainly focus on BBH merging samples with SNR$>8$ such that the false alarm rate
is below $0.02$~\citep{2016PhRvD..93l2003A}. 
Figure~\ref{fig:mc_sample_ob} show some examples of those BBHs with SNR$>8$ in different
observatories. 
We can see that in LISA and TianQin (TaiJi is also similar), many inspiralling BBHs show little evolution 
in frequency while others will evolve quickly and can be also detected in Einstein or aLIGO
band. In these space-based observatories, some of the highly-eccentric BBHs evolve differently in 
$f-h_c$ spaces from those of near circular ones. For BBHs with SNR$>8$ in 
Einstein or aLIGO band, many of them are not previously detectable in LISA/TaiJi/TianQin band. 
{ For most samples, their GW radiation can be continuously monitored by DECIGO, 
suggesting that DECIGO will be an ideal observatory to trace the inspiralling and merging of these 
highly eccentric BBH mergers.}

The number of samples with SNR$>8$ in different observatories 
can be estimated as follows. Suppose that by using Monte-Carlo method described in Section~\ref{subsec:mc_inspiral}, 
a total number of $N_{\rm mc}$ is generated and among them there 
are a number of $N_{\rm ob, mc}$ samples with SNR$>8$, then the number of samples in 
reality is given by $N_{\rm ob}=N_{\rm ob, mc}/N_{\rm mc}\times N_{\rm tot}(t_{\rm CO}<1000{\rm yr}
|t_M=5{\rm yr})$, where $N_{\rm tot}(t_{\rm CO}<1000{\rm yr}
|t_M=5{\rm yr})$ is estimated according to Equation~\ref{eq:Ninsp}.
The observable number of BBHs under other conditions is also estimated similarly by the method above.
In a five-years mission, the total number of BBHs that can be observed with SNR$>8$ in different observatories, 
or under other conditions, can be found in Figure~\ref{fig:Nob}.

If using the current aLIGO/Virgo facility and observing for $5$ yrs, we found that about $N_{\rm ob}\sim 10^2-10^3$ BBHs mergers with SNR$>8$ can be detected. This number is well consistent with the number of $1.5$ per week from the current LIGO-O2 and LIGO-O3a surveys~\citep{2021PhRvX..11b1053A} (See the red horizontal line of the top left panel in Figure~\ref{fig:Nob}). If aLIGO/Virgo can upgrade to the designed aLIGO in the future, 
the expected numbers can be about ten times higher, i.e., $10^{3}-10^{4}$ with SNR$>8$ in a five 
year's mission. The number of merging events with SNR$>8$ in { both Einstein and DECIGO can be up to 
$\sim10^5$}. Note that here the upper limit of redshift is $z<2$. If we consider 
higher redshift samples, the number of observable samples for { both Einstein and DECIGO} is expected much higher. 

The total number of SNR$>8$ events in LISA/TaiJi
band can be up to $\sim200/300$. The number become $\sim10^3$ if requiring SNR$>5$. 
For TianQin, the numbers of BBHs with SNR$>8$ (SNR$>5$) is about $1-50$ ($10\sim200$), mainly due to their high noise level.

Some of the inspiralling BBHs in LISA/TianQin/TaiJi have a short time to coalescence (e.g.,$t_{\rm CO}<10$yr), and 
them will further evolve into and merging in Einstein/aLIGO band. 
These samples are interesting for multiple band observations of the same inspiralling event. 
However, because some of the BBHs have very significant eccentricities, not all of those that enter
into aLIGO or Einstein band have previously detectable in LISA/TaiJi/TianQin band. 
For example, the characteristic strain amplitude of many of these BBHs rise above
$10^{-21\sim-22}$ at $f\sim0.1-1$Hz, which is far below the sensitivity of LISA/TaiJi. 
Only a fraction ($\sim 20\%$) of the BBHs inspiralling in LISA/TaiJi/TianQin band can enter into the 
aLIGO(design) band with SNR$>8$.  
The total number of BBHs that enter into both bands of LISA-aLIGO, TaiJi-aLIGO, or TianQin-aLIGO
can be up to $\sim20$, $\sim50$, or $\sim50$ in a five years' mission, respectively.

\section{The entering eccentricity of BBHs in different observatories}
\label{sec:ob_ecc}

\begin{figure*}
\center
\includegraphics[scale=0.7]{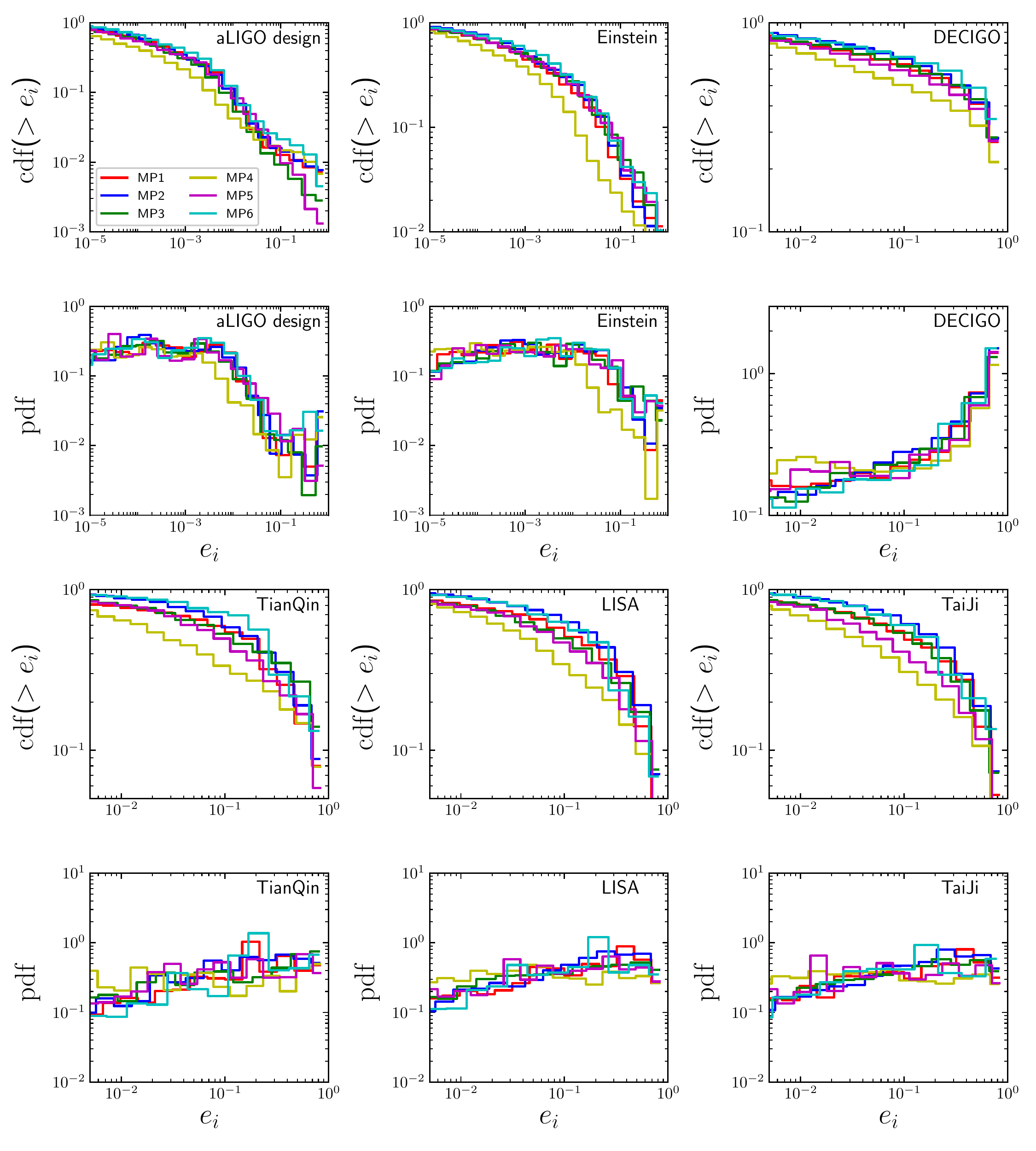}
\caption{The cumulative distribution function (cdf) or probability distribution function 
(pdf) of the eccentricity distribution of inspiralling/merging BBHs with SNR$>8$ in different observatories
and in different models.}
\label{fig:fecc}
\end{figure*}

\begin{figure*}
\center
\includegraphics[scale=0.7]{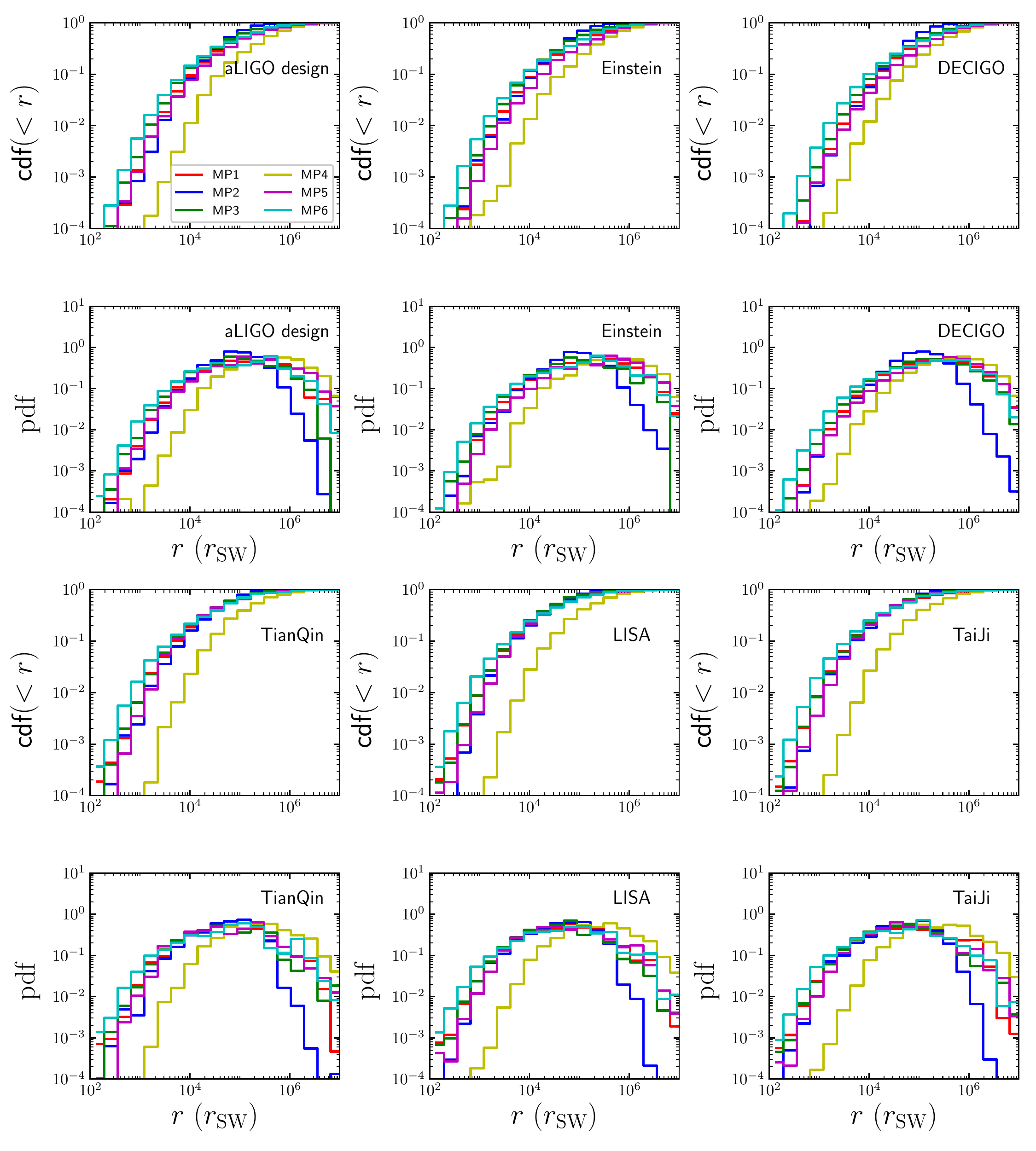}
\caption{The cdf (top panels) or pdf (bottom panels) 
of the distance of the inspiralling/merging BBHs from the central MBHs (in unit of $r_{\rm SW}$), 
which can be detected in GW observatories with SNR$>8$. }
\label{fig:r_sts_all}
\end{figure*}

The eccentricity of BBH is one of the most important features that can be used to distinguish 
different formation scenario of BBHs~\citep{Wen03,Hoang18,Antonini12,Zhang19,Seto16, 
2016MNRAS.462.2177K,2016PhRvD..94f4020N, 2016ApJ...830L..18B}. 
Here we define the ``entering eccentricity'' (denoted by $e_i$)  of a BBH in a given observatory as follows.
At the start of mission, if the peak of characteristic strain $h_c$ is below the noise level, 
then $e_i$ is the inner eccentricity ($e_1$) of BBH when $h_c$ rise and cross (for the first time) the noise level of an observatory. If initially $h_c$ is above the noise level and the GW frequency $f$ is within 
the band of observation, then $e_i=e_1$ when the mission starts ($t_s=0$).
Here we investigate the entering eccentricity ($e_i$) for samples that is with SNR $\rho>8$.
The fraction and the number of eccentric insprialling/merging 
BBHs with $e_i>0.1$ and $\rho>8$ for different observatories are shown in Table~\ref{tab:model_sts} and 
in Figure~\ref{fig:Nob}, respectively. The distribution of the entering eccentricity in 
different observatories can be found in Figure~\ref{fig:fecc}.

We find that the pdf of $e_i$ is nearly uniformly distributed between $\sim 0.1-1$ (in log scale) for LISA/TaiJi/TianQin. 
{ For DECIGO, however, the pdf of $e_i$ peaks around $\sim0.9$, as it can cover
the rising phase of those extremely eccentric BBHs.
For these space-based observatories, the fraction of eccentric events ($e_i>0.1$) is 
$\sim 30-70\%$.}
In a five year's mission, the expected number of samples with SNR$>8$ and $e_i>0.1$ 
are up to { $\sim10^5$,} $\sim100$, $\sim200$ and $\sim30$ for { DECIGO,} LISA, 
TaiJi and TianQin, respectively. 

For the ground-based observatories, the peak of pdf is much smaller. For example, 
the peak of pdf is around $e_i\sim 10^{-3}$ for aLIGO band, $\sim 10^{-2}$ for Einstein band. 
The fraction of $e_i>0.1$ is low for aLIGO(O2)/ aLIGO(design) ($\sim 1-2\%$) and Einstein ($\sim2-5\%$) 
among all the detectable events, simply because their observational windows are at high frequency. 
The expected number of events with SNR$>8$ and $e_i>0.1$
are about $1-10$ for aLIGO(O2), $10-100$ for aLIGO(design) and $\sim10^3$ for Einstein. Thus 
they should not be rare events. 

In some cases, the eccentricities of these events in aLIGO/Einstein band are quite significant, 
e.g., $e_i>0.3$, such that they rise above the noise level at GW frequency of about tens of Hz. 
For example, In Figure~\ref{fig:mc_sample_all}, the peak of GW radiation of one highly eccentric event
rises above the noise level of aLIGO directly at $f\sim20$Hz with $e_i\sim0.5$. 
Similarly in the bottom right panel of Figure~\ref{fig:mc_sample_ob}, 
one high eccentricity events rise above the noise level of Einstein at $f\sim20$Hz. 
These BBHs will merge very quickly without passing through LISA band or even 
deci-Hz band, and can only be detected in aLIGO or Einstein band. 
These BBHs provide a good explanation for 
GW 190521 observed by aLIGO/Virgo, which may have a very high eccentricity
~\citep{2020arXiv200905461G, 2020ApJ...903L...5R, 2020PhRvL.125j1102A}.

These results suggest that the eccentric inspiralling events are not only common in 
 space-based GW observatories, but also likely to be detected in ground-based observatories.
Thus it is necessary to apply eccentric waveform templates~\citep[e.g.][]{hinderer17,Seto16,2020PhRvD.101d4049L} 
to search these eccentric events for both ground and space observatories.

\begin{table*}
\center
\caption{Models and results}
\begin{tabular}{lccccccccccccc}\hline
Model  & \multicolumn{7}{c}{$\rho>8$ \& $e_{i}>0.1$  } \\
\cline{2-7}
& \begin{tabular}{@{}c@{}}aLIGO/Virgo\\(O2)\end{tabular} 
& \begin{tabular}{@{}c@{}}aLIGO\\(design)\end{tabular} & Einstein & DECIGO & TianQin  & LISA & TaiJi        \\
\hline 
MP1   &$0.7\%$    &$0.7\%$    &$3.0\%$     &$63\%$      &$46\%$     &$51\%$     &$50\%$ \\
MP2   &$1.0\%$    &$1.2\%$    &$3.4\%$     &$66\%$      &$54\%$     &$61\%$     &$59\%$ \\
MP3   &$0.5\%$    &$0.8\%$    &$4.4\%$     &$62\%$      &$54\%$     &$48\%$     &$46\%$ \\
MP4   &$0.9\%$    &$1.5\%$    &$1.8\%$     &$51\%$      &$39\%$     &$31\%$     &$31\%$ \\
MP5   &$2.3\%$    &$1.9\%$    &$4.3\%$     &$59\%$      &$51\%$     &$47\%$     &$46\%$ \\
MP6   &$1.6\%$    &$3.5\%$    &$4.2\%$     &$68\%$      &$57\%$     &$64\%$     &$57\%$ \\
\hline
%& 0.99 & 45$\arcdeg$ & 180$\arcdeg$  \\ \hline
%
\end{tabular}

\begin{tabular}{lccccccccccccc}\hline
Model  & \multicolumn{7}{c}{ $\rho>8$ \& $r<10^3r_{\rm SW}$ } \\
\cline{2-7}
&  \begin{tabular}{@{}c@{}}aLIGO/Virgo\\(O2)\end{tabular} & \begin{tabular}{@{}c@{}}aLIGO\\(design)\end{tabular}
& Einstein & DECIGO & TianQin  & LISA & TaiJi        \\
\hline 
MP1  &$0.10\%$   &$0.14\%$   &$0.11\%$  &   $0.05\%$    &$0.32\%$   &$0.47\%$   &$0.49\%$           \\
MP2   &$0\%$    &$0.1\%$   &$0.20\%$    &   $0.06\%$   &$0.18\%$   &$0.19\%$   &$0.19\%$           \\
MP3  &$0.15\%$   &$0.21\%$   &$0.15\%$  &   $0.1\%$   &$0.50\%$   &$0.55\%$   &$0.63\%$           \\
MP4  &$0\%$      &$0.01\%$    &$0.01\%$ &   $0.08\%$       &$0\%$     &$0.01\%$       &$0.01\%$           \\
MP5   &$0.1\%$    &$0.1\%$    &$0.1\%$  &   $0.06\%$    &$0.22\%$   &$0.28\%$   &$0.30\%$           \\
MP6  &$0.42\%$   &$0.49\%$   &$0.42\%$  &   $0.3\%$   &$1.50\%$   &$1.69\%$   &$1.58\%$           \\
\hline
\end{tabular}

\begin{tabular}{lccccccccccccc}\hline
Model  & \multicolumn{7}{c}{$\rho>8$ \& $\rho_{\delta \Psi}>8~\&~e_i<0.4$ } \\
\cline{2-7}
&  \begin{tabular}{@{}c@{}}aLIGO/Virgo\\(O2)\end{tabular} & \begin{tabular}{@{}c@{}}aLIGO\\(design)\end{tabular}
& Einstein &DECIGO & TianQin  & LISA & TaiJi        \\
\hline 
MP1   &$0\%$      &$0\%$      &$10^{-4}$         &$ 30\%$    &$68\%$     &$46\%$     &$54\%$           \\
MP2     &$0\%$    &$0\%$      &$4\times10^{-5}$  &$ 31\%$    &$64\%$     &$44\%$     &$51\%$           \\
MP3     &$0\%$    &$0\%$      &$8\times10^{-5}$  &$ 31\%$    &$52\%$     &$39\%$     &$45\%$           \\
MP4   &$0\%$    &$0\%$      &$10^{-5}$           &$ 26\%$    &$50\%$     &$37\%$     &$41\%$           \\
MP5   &$0\%$      &$0\%$      &$3\times10^{-5}$  &$ 29\%$    &$56\%$     &$42\%$     &$47\%$           \\
MP6   &$0\%$      &$0\%$      &$10^{-4}$         &$ 25\%$    &$47\%$     &$46\%$     &$53\%$           \\

\hline
%& 0.99 & 45$\arcdeg$ & 180$\arcdeg$  \\ \hline
%
\end{tabular}
\tablecomments{ 
Percentages of number of samples with 
parameters larger or smaller than some value of interest over the number of all observable samples (SNR$>8$)
in different observatories.
Here $e_i$ is the entering eccentricity of the BBH (See texts in Section~\ref{sec:ob_ecc} for its definition), 
$r$ is the distance of the BBH to the central MBH in unit of Schwarzschild radius, 
$\rho_{\delta \Psi}$ is the SNR for phase drifts given by Equation~\ref{eq:snr_phase_drift}. 
}
\label{tab:model_sts}
\end{table*}

\section{The inspiralling/merging BBHs close to the MBH}
\label{sec:relativity_BBHs}
\begin{figure}
\center
\includegraphics[scale=0.7]{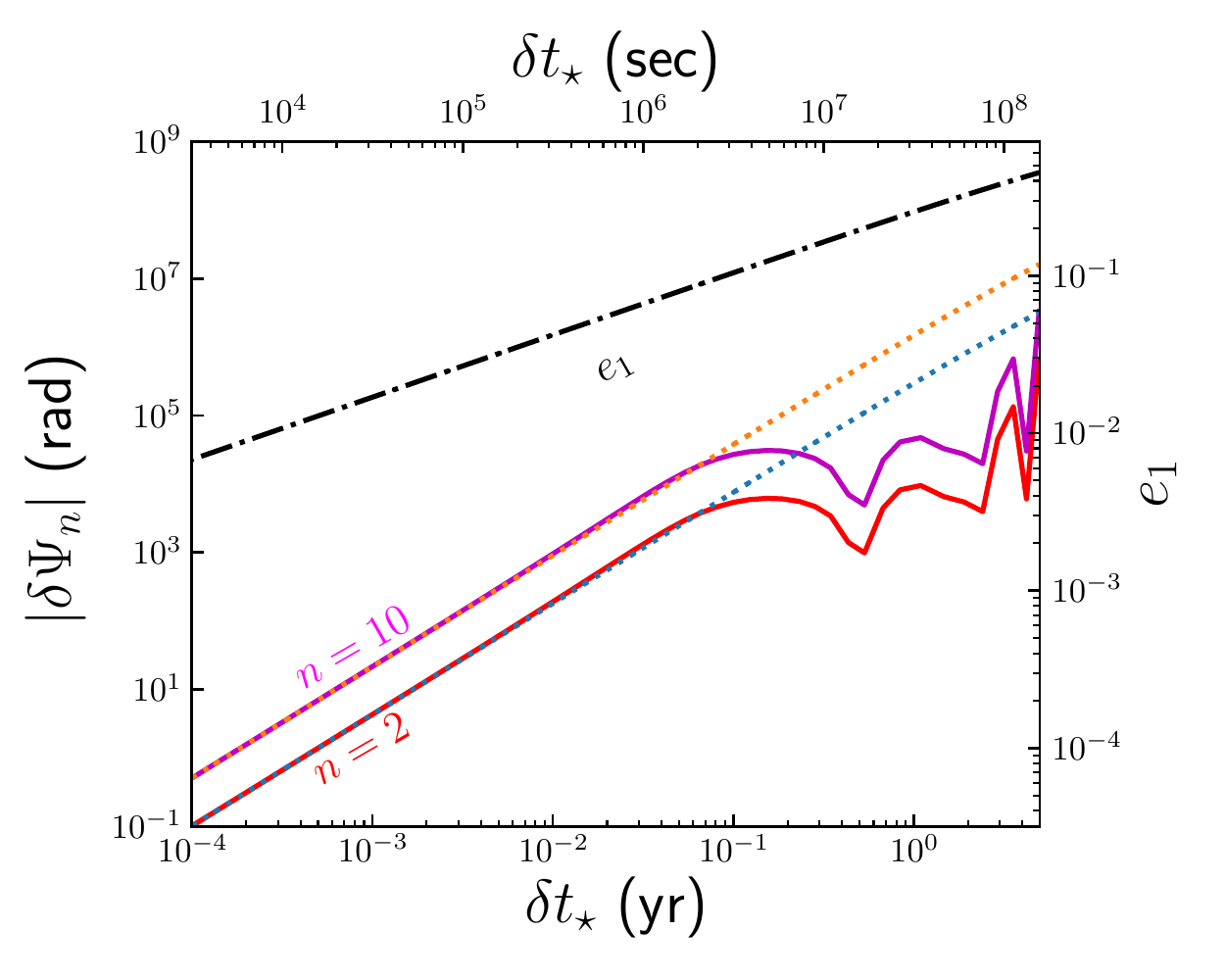}
\caption{The increase of Fourier phase drift of an example BBH rotating around MBH as a function of time to coalescence
$\delta t_\star$. Here $\bh=10^6\msun$, the BBH is with $m_A=m_B=10\msun$, $a_2=100\AU$
and $e_2=0.6$, where $a_2$ and $e_2$ is the SMA and eccentricity of the outer orbit of 
BBH, respectively. The BBH coalesces with $e_{\rm iso}=5\times10^{-5}$. 
The dash-dot line show the evolution of the inner eccentricity of the BBH ($e_1$).
The dotted lines are prediction from Equation~\ref{eq:deltapsismall}; the solid lines
are results from Equation~\ref{eq:deltapsin0}.}
\label{fig:phase_drift}
\end{figure}

\begin{figure*}
\center
\includegraphics[scale=0.7]{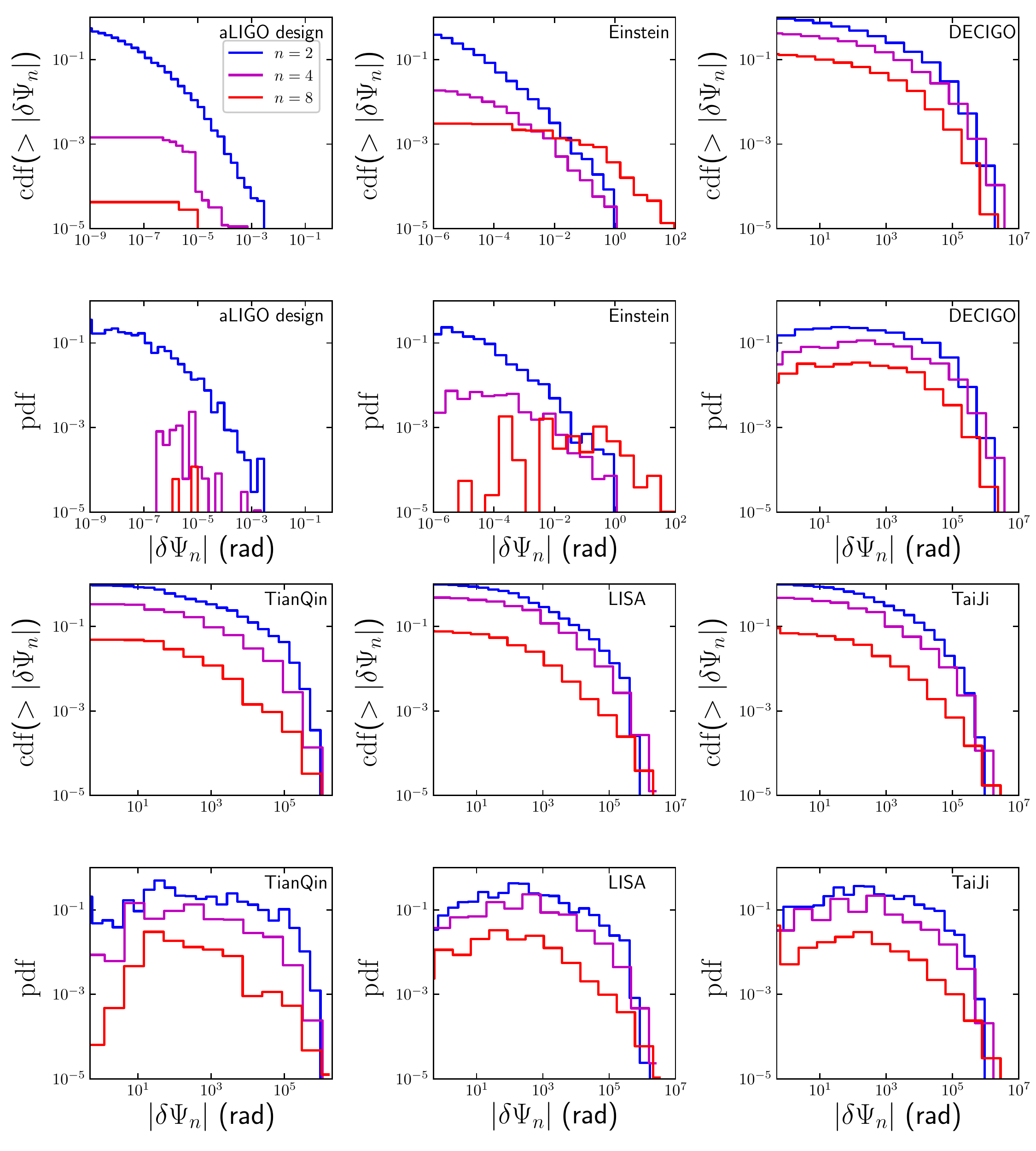}
\caption{The cdf (top panels) or pdf (bottom panels) 
of the Doppler drift $\delta \Psi_n$ in Fourier phases of 
the inspiralling or merging BBHs in different observatories 
(only for samples with $e_i<0.4$ and SNR $\rho>8$) in model MP1. Here $n$ is the order of harmonics.
Only $\delta \Psi_n$ of samples with strain $h_{c,n}$ of the $n$-th harmonic above the sensitivity 
of the observatory are  considered in this plot. The results for other models are similar. }
\label{fig:dpsi_sts_all}
\end{figure*}

As shown in Section~\ref{subsec:evl_merging_BBHs}, there is a fraction of BBHs that can inspiral/merge 
very close to the MBH (See Table~\ref{tab:model_frac}). 
The inner most orbit of BBHs is limited by the tidal radius, thus the periapsis of the outer 
orbit of BBH should satisfy:
\be\ba
r_{p,\rm min}&\ga a_{1}\left(\frac{3\bh}{m_A+m_B}\right)^{1/3}\\
\ga & 266  r_{\rm SW}\times\frac{a_{1}}{0.1~\rm AU}\left(\frac{\bh}{10^6\msun}\right)^{-2/3}
\left(\frac{m_A+m_B}{20\msun}\right)^{-1/3}
\label{eq:rp_min}
\ea\ee
where $a_{1}$ is the SMA of the inner orbit of BBH. If 
$a_1=0.1\rm AU$, $r_{p,\rm min}$ varies between $10^3-10r_{\rm SW}$ 
for MBHs with masses vary from $10^5\msun$ to $10^8\msun$. 

By using the samples obtained in 
Section~\ref{subsec:mc_inspiral}, we can estimate the distributions and the 
number of inspiralling/merging BBHs with SNR$>8$ and at distance $r<10^3r_{\rm SW}$ in different observatories. 
The results are shown in Figure~\ref{fig:Nob} and ~\ref{fig:r_sts_all} and Table~\ref{tab:model_sts}.

Figure~\ref{fig:r_sts_all} show the pdf of the distance of the inspiralling/merging 
BBHs to the MBH in different observatories. The distribution of BBHs peaks around $10^5r_{\rm SW}$
where $r_{\rm SW}$ is the Schwarzschild radius, almost independent with the observatories.
The fraction of BBHs within some given radius $r$
decreases rapidly with $r$. From Figure~\ref{fig:r_sts_all} and Table~\ref{tab:model_sts},
we can see that the probability of $r<10^3r_{\rm SW}$ is about $0.1\%-2\%$. 
Thus for aLIGO/Virgo(O2), aLIGO(design), or Einstein, in a five years mission, there are about $\sim1$, 
$1\sim10$ and $10-100$ of them can be detected, respectively (See Figure~\ref{fig:Nob}).
{ For DECIGO, the expected number can be $10-1000$, however, for 
LISA/TaiJi, it is very marginally ($\sim1$)}. TianQin can not observe 
such samples as their expected number are smaller than one. 
We note that the space observatories are more powerful in revealing the relativistic 
effects of the samples that is very close to the MBH, as their observation 
time is much longer than those in aLIGO and Einstein. 

Due to the existence of a MBH in the vicinity of the inspiralling/merging BBH, the GW 
radiation will show some unique properties, which is quite different from those BBHs in 
other environments (e.g., the BBHs in the field regions or globular clusters). 
For example, if the inspiralling/merging BBHs are sufficiently
close to the MBH, they will experience significant accelerations 
or relativistic effects including gravitational redshift, lensing, and relativistic orbital
precession. These effects modify the observed gravitational waveforms in a way very similar to those of the pulsar timing observation, which could cause de-phase of the wave and thus possibly measurable 
in the current or future GW observatories~\citep[e.g.,][]{2017ApJ...834..200M,
Inayoshi17}. These BBHs provide unique probes for testing relativity in the vicinity of MBHs, and 
their unique features in waveforms can be used to distinguish the BBHs merging 
around MBHs in galactic nucleus from those in other environments. 

As a first study, here we investigate and discuss only the Doppler drift effects due to the 
accelerations of the BBHs around a MBH. These effects can be easily analyzed 
by using some analytical methods if the orbits of the BBHs are near circular.
However, it may be more difficult for eccentric BBHs, as their evolution
in the Fourier domain is more complex than those 
of circular ones~\citep{2016PhRvD..93f4031T,2009PhRvD..80h4001Y,2016PhRvD..94f4020N,2018CQGra..35w5006M}. 
For simplicity, here we only estimate the Doppler drift 
effects of BBHs with entering eccentricity $e_i<0.4$, by using the 
Fourier phase formalism from~\citet{2009PhRvD..80h4001Y} which ignore other PN orders of the orbit
(expect the GW orbital decay). 
The Doppler drift effects for higher eccentric events or including other PN orders 
will defer to future studies.

According to Figure~\ref{fig:fecc}, the eccentricities of BBHs are usually 
quite low in aLIGO and Einstein observations, with $\ga 98\%$ of $e_i<0.4$, and 
the fraction of BBHs with $e_i<0.4$ is { $60\sim80\%$ in DECIGO/LISA/TianQin/TaiJi band}.
Our estimation of the Doppler drift should cover a majority of all the detectable events.

Details of the derivation of the Fourier phase drift of BBHs due to its Keplerian motion around MBH can be found in Appendix~\ref{apx:fourdrift}. To estimate the phase drift given the order of harmonic $n$ 
we need to select a time of reference. It is defined as the 
time at the end of the mission, or the time when the characteristic strain of the BBH $h_{c,n}$ 
of the $n$-th harmonic moves just below the detection sensitivity.
For aLIGO/Einstein and $n=2$, it is usually the time of coalescence. 
If $h_{c,n}$ of the $n$-th 
harmonic is below the sensitivity during the whole mission time, we consider that it is not detectable.

In the case when the eccentricity of BBH $e_1$ is small and the coalescence time is the time of reference, 
the phase drift of the $n$-th harmonic 
can be estimated by Equation~\ref{eq:deltapsismall} (up to orders of $e_i^2$)
\be\ba
|&\delta \Psi_n |=0.102\times M_6^{-1} r_{4}^{-2} f_{1{\rm Hz}}^{-13/3}\mathcal{M}_{14}^{-10/3}
\left(\frac{n}{2}\right)^{16/3}\sin I_2\\
&\times\mathscr{P}(E',e_2, \omega_2)\left(1-\frac{314}{43}e_i^2\chi^{-19/9}\right)
\label{eq:deltapsismall2}
\ea\ee
where $e_i$ is the entering eccentricity of the BBH, $\chi=f/f_i$ and $f_i$ is the  
GW frequency of the BBH when $e_1=e_i$, $M_6=\bh/10^6\msun$, $\mathcal{M}_{14}=\mathcal{M}/14\msun$, 
$r_{4}=r/10^4{~r_{\rm SW}}$ and $f_{1{\rm Hz}}=f/1{\rm~Hz}$.
$\mathscr{P}\la 1$ is a factor determined by the line of sight position of BBH, given by
\be
\mathscr{P}=
\frac{\sin \omega_2(\cos E'-e_2)+(1-e_2^2)^{1/2}\cos \omega_2\sin E'}{1-e_2\cos E'}
\ee
where $E'$ is the eccentric anomaly at the reference position. $e_2$, $I_2$ and $\omega_2$ is the eccentricity, 
inclination and the argument of periapsis of the outer orbit of the BBH around MBH.

We can see that at a given time, $\delta \Psi_n\propto (f/n)^{-13/3}n\propto n$, 
thus it will be easier to observe 
phase drift effects in the high order harmonics for eccentric BBHs.
However, note that $h_{c,n}$ of the $n$-th harmonic has to be above the noise level of an 
observatory such that it can measured.
An example of the phase drift of a BBH is shown in Figure~\ref{fig:phase_drift}. 

The drift of the Fourier phase of a BBH due to acceleration can be observed 
only if the perturbations on the strain amplitudes are large enough. 
The signal to noise ratio can be estimated by ~\citep{2011PhRvD..84b4032K}
\be
\rho^2_{\delta \Psi}=
2\sum^{n_{\rm max}}_{n=1}\int^{f_{\rm max}}_{f_{\rm min}}
\frac{h_{c,n}^2(f)(1-\cos \delta \Psi_n)}{fS_n(f)}\frac{df}{f}
\label{eq:snr_phase_drift}
\ee

To detect such perturbations in observatories, it is required that at least one of the 
$n$-th harmonics of $h_{c, n}$ ($n=1, 2, \cdots$) is sufficiently above the noise level, 
and that the corresponding $|\delta \Psi_n|$ is also sufficiently larger than unity.
For aLIGO, the eccentricities of most BBHs are quite low, thus the most important order is $n=2$. 
According to Equation~\ref{eq:deltapsismall2}
we have $f_1=10$ and $|\delta \Psi_2|\sim 5\times10^{-4}$ even if $r_4\sim 0.1$,  thus it is not 
possible to observe Doppler drift in aLIGO band. 
Such a small drift is a result of the extremely short time the BBHs spent in aLIGO band 
(typically in orders of $1-10$ seconds) respect to the orbital time of BBH around MBH 
($\ga$days if $a_2\ga10^3r_{\rm SW}$). 
Figure~\ref{fig:dpsi_sts_all} show the distributions of $\delta \Psi_n$ for different orders of $n$ for 
those BBHs with $\rho>8$. We can see that for aLIGO the maximum possible value of $|\delta \Psi_2|$ is about 
$10^{-3}$, and we do not expect to see any Doppler drift effects in aLIGO (See also Figure~\ref{fig:Nob}).

For Einstein the lookback time of BBHs coalescence can be about a few hours and the 
eccentricities of BBHs are slightly higher than aLIGO. 
It is interesting to see that the large phase drift come from those harmonics of high order 
(See Figure~\ref{fig:dpsi_sts_all}), as for many highly-eccentric BBHs, 
only high order harmonics of strains appear earlier in
Einstein band, instead of the low order ones (See the bottom panels 
of Figure~\ref{fig:harmonic_full_ob}) at a given time. 
Thus, according to Equation~\ref{eq:deltapsismall2}
the Doppler drift can be observable if $f_1\sim1$, $r_4\sim 0.1$ and $n>2$, however, the
probability of $r_4\la 0.1$ is about $10^{-3}$ as shown in Figure~\ref{fig:r_sts_all}.
Nevertheless, we find that the expected number of samples with $\rho>8$ and $\rho_{\delta\Psi}>8$ 
in Einstein telescope can be up to $\sim10$ (See Figure~\ref{fig:Nob}). 

For the space observatories, the drift of phase in a $5$-yr's mission is so large that 
the approximation given by Equation~\ref{eq:deltapsismall2} is no longer valid, and we need to use explicitly 
the Equation~\ref{eq:deltapsin0} for accurate estimation. As shown in Figure~\ref{fig:dpsi_sts_all}, 
in { DECIGO/}LISA/TaiJi/TianQin band, the distributions of $\delta \Psi_n$ for $n=2$, $n=4$ and $n=8$ all peak 
around $10-10^3$. Most of the BBHs are with phase drift contribute 
from $n=2$, which is larger than those of higher order ones, such as $n=4$ and $n=8$.
This is mainly because { we focus on samples with $e_i<0.4$, and thus} 
many of the high order harmonics of GW are below the sensitivity level of 
space observatories (See Top-right panel of Figure~\ref{fig:harmonic_full_ob}).
The observable numbers of phase drift with $\rho_{\delta \Psi}>8$ and $\rho>8$ 
can be up to {$10^4-10^5$ for DECIGO,} $\sim30$ for TianQin and $\sim100$ for LISA/TaiJi, which is about 
$\sim30-80\%$ of all detectable samples, see Table~\ref{tab:model_sts}. 

Note that the above estimations are for samples with entering 
eccentricity $e_i<0.4$. Thus, for space telescopes the actual number of samples with
detectable Doppler drift should be higher{, especially for DECIGO}. 

\section{Gravitational wave backgrounds of BBHs}
\label{sec:GWBK}
\begin{figure}
\center
\includegraphics[scale=0.7]{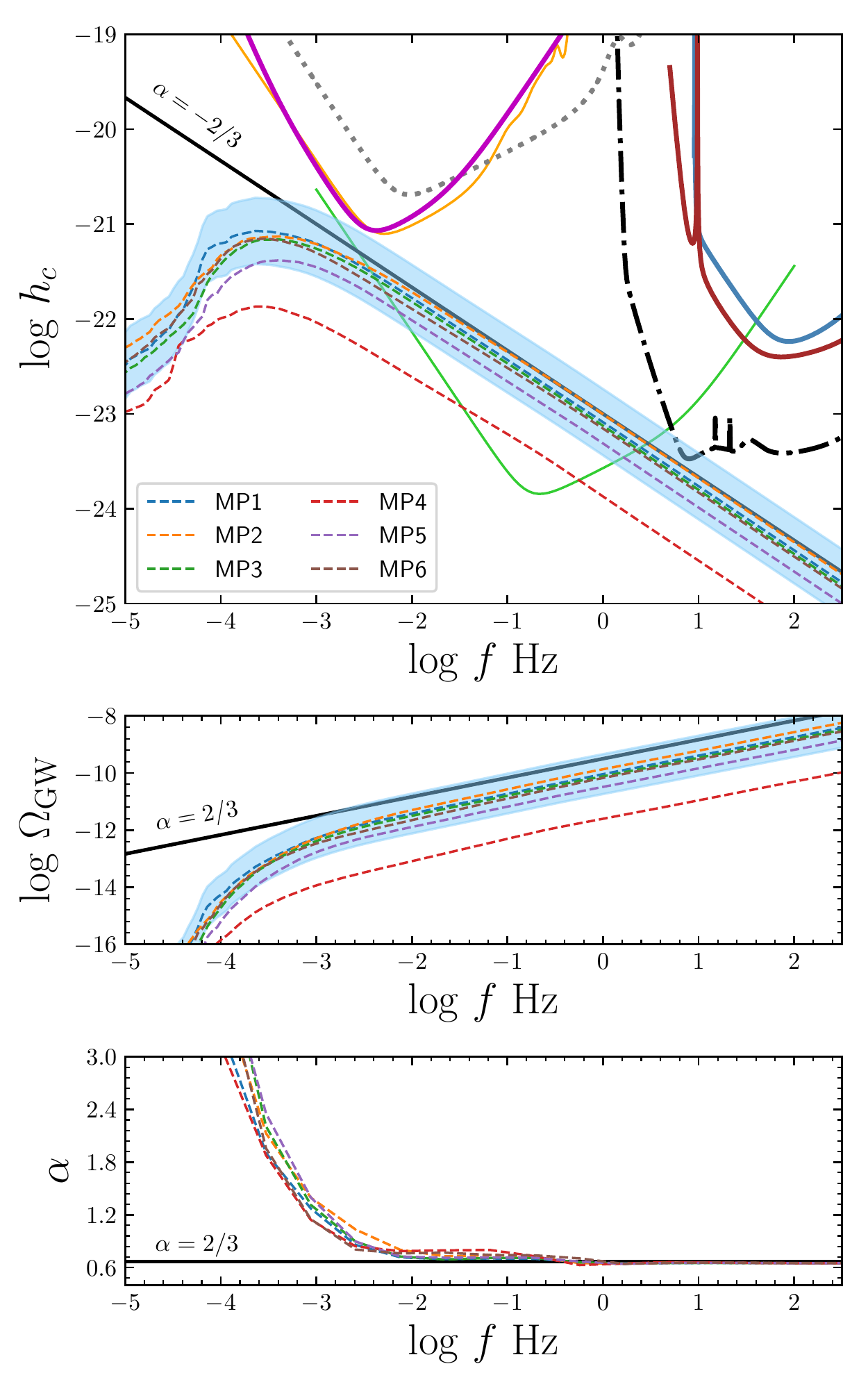}
\caption{GWB in different models. Top panel: The characteritic strain amplitude of GWB from different 
models. The shaded region show the uncertainties in model MP1, due to the uncertainties in event rates. The uncertainties 
is similar for other models and thus not showed here for clarity of the figure. 
The black solid line show a reference slope of $\alpha=-2/3$; Middle panel: Similar to top panel but for 
the energy density of GWB from different models. The black solid line show a reference slope of 
$\alpha=2/3$; Bottom panel: The powerlaw slope of $\Omega_{\rm GW}$ 
in different models as a function of frequency. 
}
\label{fig:gwbk}
\end{figure}

The total number of inspiralling BBHs at any moment can be up to billions (See the estimation at the
end of Section~\ref{subsubsec:num_insp_merge}). They appear in a GW observatory as a background noise 
as the majorities of them are weak and slowly evolving BBHs. As many of the BBHs inspiralling/merging around MBH 
are with high eccentricity, the gravitational wave background (GWB) contributed by these objects are expected
different from those of circular inspiralling/merging BBHs~\citep[e.g.,][]{2020MNRAS.500.1421Z}. Here we investigate the GWB by using the BBH samples obtained from {{\texttt{ GNC}}} simulations.

The GWB can be estimated by~\citep{2001astro.ph..8028P}
\be\ba
&h_{c,bkg}^2(f)=\frac{4G^{5/3}}{3\pi^{1/3} c^2f^{4/3}}
\iiint de_0 d\mathcal{M}dz \times\\
& \frac{d^3N(\mathcal{M},e_0,z)}{d\mathcal{M}de_0 dz} \frac{ \mathcal{M}^{5/3}}{(1+z)^{1/3}}
\sum_{n=1}^\infty\left(\frac{2}{n}\right)^{2/3}\frac{g(n,e)}{F(e)}\\
\label{eq:hcbkg_int}
\ea\ee
where $g(n,e)$ and $F(e)$ is given by Equation~\ref{eq:gne} and~\ref{eq:fe}, respectively.
$d^3N(\mathcal{M},e_0,z)/d\mathcal{M}de_0dz$ is the number distribution function of BBHs.
The eccentricity $e$ in the above equation is determined by the following relation
for a given value of $e_0$, $n$ (the order of harmonics) and $f$:
\be
\frac{f}{f_{0}}=\left[\frac{1-e_0^2}{1-e^2}
\left(\frac{e}{e_0}\right)^{12/19}\left(\frac{1+121e^2/304}{1+121e_0^2/304}\right)^{870/2299}\right]^{-3/2}
\ee
where $f_{0}=nF_{\rm orb}$, and $F_{\rm orb}$ is the initial orbital frequency 
given the initial eccentricity $e_0$.  

We can obtain individual BBH samples from the {{\texttt{ GNC}}} rather than its distribution, thus 
we can use the Monte-Carlo integration to estimate the strain amplitude of the GWB, 
which is much more simple and straight foward. For more details see Appendix~\ref{apx:GWBK}.

The strain of GWB for different models can be found in Figure~\ref{fig:gwbk}.
We found that the characteristic strain amplitude of GW backgrounds follows well
a powerlaw profile of $h_{c,bkg}\propto f^{-2/3}$ when $f>1$mHz, similar to those 
from circular BBHs~\citep{2001astro.ph..8028P}. However, the slope changes below $f<1$mHz as the 
eccentric inspiralling BBHs start to dominate the GWB at those low frequency regions, 
which is also consistent with those from~\citet{2020MNRAS.500.1421Z}. 
The energy density of the GWB, i.e., $\Omega_{\rm GW}=2\pi^2h_{c,bkg}^2(f)f^2/(3H_0^2)$,
for different models are shown in Figure~\ref{fig:gwbk}. Similarly, we find that the energy density 
$\Omega_{\rm GW} \propto f^{2/3}$ if $f\ga1$mHz. The power-law slope changes to
$\sim 1.2$ at $f\sim1$mHz and higher if $f<1$mHz. As LISA can put constraints on the GWB
above $\Omega_{\rm GW}>10^{-14}$ at $f\sim 1$mHz~\citep[See Figure 3 of][]{Abbott17a}, 
we expect that LISA/TaiJi{/DECIGO} can be used to distinguish different merging channels by measuring the profile of GWB.

\section{discussion}
\label{sec:discussion}
There are some unique phenomena proposed for BBHs merging around MBH. For example, the b-EMRIs~\citep{2018CmPhy...1...53C}, Doppler abbreviation~\citep{2020PhRvD.101h3028T}, relativistic 
effects~\citep{2017ApJ...834..200M}, lensing effects\citep{2020PhRvD.101h3031D,2013ApJ...763..122K}, and etc. To observe these effects, the position of the inspiralling/merging BBHs should be very close to the MBH. We find 
that there are mainly two factors that affect the closest distance between a BBH and MBH. 
The first one is the size of the loss cone region of the MBH, which is due to the tidal force of the MBH and 
it puts a very solid boundary of the BBHs. Thus, almost independent of the initial conditions of the models 
(expect model MP4, see below), the probability of BBHs merging at distance $r<10^3r_{\rm SW}$ is always about
$10^{-3}-10^{-2}$. The second factor is the mass of the background objects. If the background object is massive (e.g., $m_\star=10\msun$ in model MP4), the encounters of BBHs with them have a high probability of ionization and 
 they will likely be destroyed before moving into the inner most region.

Comparing models of MP1 and MP2, we find that placing the birth position of BBHs into more inner regions can 
increase simultaneously the fraction of BBH being merged, tidally disrupted, and ionized (such that BBHs remain integrity is reduced). This will result in a higher merging rate and the predicted total number of samples detectable 
($\rho>8$) in different observatories. However, the fraction of samples in the 
vicinity of MBH are almost unchanged, mainly due to the solid boundary 
of the loss cone set by the MBH. 

Comparing the results of MP3 to other models, we find that if initially the BBHs are born with zero
eccentricity, the merging rates, distribution of the merging position of BBHs and other properties of BBHs
are not significantly affected. This is mainly because the information of the initial value of the eccentricity 
will be lost after experiencing large number of multiple dynamical process, including the KL oscillation, encounters 
with the background objects or with the MBH. The distribution of eccentricities of BBHs in an equilibrium state is 
then shaped by these dynamical evolution rather than the initial condition of the eccentricities.

Comparing models of MP6 to other models, we find that if the background components are mixed with single 
black holes ($10\msun$) and stars ($1\msun$), the dynamical evolution of BBHs is similar to those 
assuming background stars with $1\msun$. This is mainly because the single BHs always contribute a small 
fraction of the numbers ($\sim10^{-3}$) and masses ($\sim10^{-2}$) of the cluster, 
expect those of the most inner regions.
For a Milky way size MBH, the enclosed mass of BHs surpass those of the single stars within radius of
about $10^3$ AU~\citep{Alexander09}, thus the majorities of the cluster are not affected by the existence 
of these BHs. The overall effect is a slight increase of the ionization rates of the BHs (from $\sim 1-7\%$
to $5-11\%$).

{Assuming that the IMF of stars formed in galactic nuclei follows a top-heavy one, 
effective mass segregation due to relaxation leads to a mass function that appears to be between 
the ``LIGO\_BK'' of MP1 and Salpeter-like IMF of MP5. To explore the effects, we run an additional simulation 
of model that is similar to MP5 but with $\propto m_A^{-1.8}$. We find that both the event rates
and the predicted number of samples in different observatories are 
between those predicted from the two above models. This suggests that the two model would provide an 
upper and lower bound of the predicted numbers of the merging BBHs, respectively. 
}

Here in this work we limit our samples with $z<2$. For higher redshift the structure of the nuclear star cluster 
around the MBH may be different from those in local universe, thus the results may not be reliable 
if we expanding to higher redshift. However, Einstein telescope can observe samples with $z>2$, and 
our predicted numbers of Einstein is only a lower limit of their total sample sizes. The samples from Einstein
telescope in the future can help to study the evolution of BBHs in galactic nuclei in high redshift.

\section{{The planned future updates of the Monte-Carlo code}}
\label{sec:update}
{
It is very important to study the hierarchical formation of BBHs and the distinct 
signatures between the first and multiple generation of 
BBH mergers~\citep[e.g., see a recent review in][]{2021NatAs...5..749G}.
It is suggested that GW190521 might come from second-generation mergers 
due to its large component masses inside the mass gap~\citep{2020ApJ...900L..13A}, 
supporting their dynamical formation hypothesis. In the meanwhile, investigating 
presented features of EMRIs together with those BBH mergers in galactic nuclei 
is also important to understand the formation and 
evolution of MBH in near-by universe~\citep[See a review in][]{2018LRR....21....4A}.}

{
However, due to some limitations of the current \texttt{GNC} code, we are unable to 
provide reasonable estimations for these GW events. For example, if the exchange event of a BBH with 
a background object happens in current \texttt{GNC} code, the simulation stops and thus the following 
evolution of the BBHs are not calculated and the sample is abandoned for simplicity. 
Also, the evolution of a BBH stops when it is found merging, and thus the possibility that 
its merged remnant merging with another BH (i.e., producing the second generation merger) are not included in the 
simulation. Another limitation is that each components of the cluster is assumed to follow a
given power law distribution (e.g. for stars $\propto r^{-7/4}$ or $\propto r^{-1.4}$) and 
the depletion of background objects due to tidal disruption (for stars) or falling into MBHs 
(for compact objects) and the exchange between the background objects with BBHs are not included 
in the simulation. Thus, the results presented in this work are all for first generation BBH 
mergers. For comprehensive and self-consistent study of GW merging events, including 
multiple generation of BBH mergers and also the EMRIs in galactic nuclei, we plan to include 
the following functionalities into \texttt{GNC} in future:
\begin{itemize}
\item Evolving multiple species background component. As these background components can be also affected by 
the tidal disruption of MBH (for stars), or EMRIs (for compact objects), the rejuvenation of them 
due to the disrupted, ionized or merged BBHs, we can update the profile of background objects 
according to these events and then obtain a self-consistent steady-state Fokker-Planck solution 
of the density profiles for the background components.
\item Follows the evolution of both the BBHs that have exchanged components with background objects, 
and the evolution of component being exchanged into background species. For example, a BBH may become 
a BH-MS binary with the exchange with a background star, and the BH component exchanged become one of
the background BH species. If a BBH exchange with those first-generation BBH mergers, 
a second generation BBH mergers may happen.
\item The dynamical evolution of individual background objects, including two-body and resonant relaxations. 
By including such evolution, we are then able to study the dynamical evolution and the properties (e.g., event rates) 
of EMRI in galactic nuclei. 
\item The spinning of the merging BBHs and the GW recoil kick velocity due to the merging of a BBH. 
The spin distribution of the hierarchical BBH mergers can possible be 
affected by the dynamically evolution and GW recoil kick may significantly affect the orbital evolution 
of the BBH mergers, if it is not kicked out from the galactic nuclei. The distribution of spin and also 
the masses of the BBH mergers provide another dimension in distinguishing different merging channels.
\item Stellar evolution, such that the composition of background components and also the compact binaries as 
a result of stellar evolution of stars (or binary stars) can be estimated self-consistently under different 
metalicities. In the meanwhile, the stellar evolution of those BH-MS binary can be traced, which
may become BBH or other black hole compact binaries in the later stage of the simulation.
\item Tidal dissipation of stellar objects. For those BH-MS binary during the Kozai-Lidov oscillation,
and those BBHs encountering with a background stars, the tidal dissipation may affect the 
evolution significantly, and thus tidal dissipation should be added into the code for self-consistency.
\item Considering various components of compact objects, such as neutron stars (NS) or white dwarfs (WD) in both 
the background species and also the compact binaries. Such that we can be able to discuss the 
evolution and merging of various types of compact binaries, especially BH-NS, BH-WD binaries or NS binaries.
\end{itemize}}

{By implementing the above features, \texttt{GNC} will become a more powerful tool to study 
the complex interplay between various types of compact objects under a number of 
dynamical effects around MBH. Many of these phenomena can be treated in a self-consistent 
way, such that we can provide accurate predictions of the hierarchical binary merging events 
(e.g., distribution of mass, spin, eccentricity, and other 
unique features) and EMRI events. These predictions can then be useful for 
distinguishing GW merging sources in galactic nuclei and those in other merging channels.
}
\section{conclusions}
\label{sec:conclusion}
Stellar binary black holes (BBHs) merging around the MBH have same unique features, 
for example, extremely high eccentricities and locating at the very vicinity of the 
MBH that cause very large accelerations of their motion. We study the properties of these 
unique features by { updated numerical methods of~\citet{Zhang19} (named \texttt{GNC})}. 
We also study the detection of these BBHs
in ground or space gravitational wave observatories, 
including aLIGO/Virgo, Einstein, { DECIGO,} LISA, TianQin and TaiJi. 

We find that $3-40\%$ of all newly formed BBHs will finally merged due to various dynamical effects. 
A significant fraction of them are merged due to encounters with the background stars ($7\%-90\%$) 
and KL effects ($6\%-70\%$). The rest of them ($1-25\%$) 
are merged after one or multiple encounters with the MBH. Some of these events 
($40\%-70\%$) can have extremely high eccentricities ($1-e_1\la 10^{-3}$), or locate very close to the MBH 
with $r<10^4 r_{\rm SW}$ (up to $10\%$) when the gravitational wave dominates the inner orbital evolution.

We find that in a five year's mission, the number of inspiralling/merging
BBHs with SNR$>8$ can be up to $\sim 10^4$ in aLIGO, $\sim 10^5$ in Einstein{/DECIGO}, 
$\sim50$ in TianQin, and $\sim 200$ in LISA/TaiJi band. 
{ Some} of them can be observed in multiple band observations.
For example, there are about $\sim 10$ samples both detectable 
in LISA/TaiJi/TianQin and aLIGO band with SNR$>8${;  Up to $\sim10^4$ samples can be both 
detectable in DECIGO and aLIGO band with SNR$>8$.}

Due to the existence of the central MBH, these BBHs have two unique characteristics: 
\begin{enumerate}
\item
Significant eccentricities. The fraction of high entering eccentricity 
$e_i>0.1$ for different observatories are: aLIGO/Virgo(O2)/aLIGO(design) ($1-3\%$), Einstein ($2-7\%$), {{ DECIGO ($50-70\%$)}}
TianQin ($30-65\%$) and LISA ($30-90\%$). { The significant  eccentricities of these events provide a possible explanation 
for the event GW 190521}. Due to very high orbital eccentricity, many of the BBHs entering into the 
aLIGO/Einstein band are not previously detectable in LISA/TaiJi/TianQin band as
the strain amplitude become significant only if gravitational wave frequency $>0.1$Hz. 
{ Thus, DECIGO become the most ideal obsevatory in probing the rising phases of these extremely 
high eccentricy GW events.} 

\item We find that $\sim0.1\%-2\%$ of the BBHs can merge in distance less than $10^3 r_{\rm SW}$. 
These samples have very large acceleration 
and may also have some significant relativistic effects. We explore the Doppler drift effects in the 
Fourier phases which can be detectable with SNR$>8$. We find that the number of them with SNR$>8$ 
are up to { $\sim 10^4-10^5$ in DECIGO}, $\sim 30$ in TianQin and $\sim10^2$ in LISA/TaiJi. 
For Einstein the expected numbers of detection can be up to $\sim10$. 
\end{enumerate}

We find that the gravitational wave background from these highly-eccentric BBH have an energy density follows a powlaw 
slope of $2/3$ if $f>1$mHz, but deviate to $\ga1.2$ at frequency $f<1$mHz. 
Future space-based telescopes, such as LISA/TaiJi{/DECIGO} can be able to detect such deviations.

The high eccentricity, the strong accelerations (and also relativistic effects) and a different profile of GWB expected 
from these sources make them distinguishable from other sources, thus interesting for future GW detection and tests of relativities.
\acknowledgements

\noindent
We thank Huang Qingguo, Lu Youjun and Zhao Yuetong for helpful discussions.
This work was supported in part by the Natural Science Foundation of Guangdong
Province under grant No. 2021A1515012373, National Natural Science Foundation of 
China under grant No. 11603083, U1731104. This work was also supported in part by
the Key Project of the National Natural Science Foundation of China under grant No. 11733010. 
X.C. is supported by NSFC grants No. 11873022 and 11991053.
LS was supported by the National Natural Science Foundation of China (11975027, 11991053, 11721303), the National SKA Program of China (2020SKA0120300), the Young Elite Scientists Sponsorship Program by the China Association for Science and Technology (2018QNRC001), and the Max Planck Partner Group Program funded by the Max Planck Society. 
The simulations in this work are performed partly in the TianHe II National
Supercomputer Center in Guangzhou.

%-----------------------------------------------------------------------

\appendix
\section{Updates to the Numerical Method}
~\label{apx:method_update}
The numerical Monte-Carlo framework previously developed in~\citet{Zhang19} have now been 
updated, with the following improvements:
\begin{enumerate}
\item We can now consider both the NR and RR diffuion processes of the BBHs under background 
objects with multiple species. For the two-body relaxation, 
such extension can be realized by noticing that the diffusion coefficients of BBHs 
are obtained by suming up the contributions of all the background particles. 
Following~\citet{1957PhRv..107....1R,Binney87}, we can show that 
if there are a number of $N_b$ kinds of background objects with mass $m_i$, 
number density function $n_i(r)=n_{i,0}(r/r_h)^{\alpha_i}$, here $i=1,\cdots, N_b$, 
where $n_{i,0}$ is the number density of the $i$-th component at position $r_h$, 
$\alpha_i$ is the index of the density profile of the $i$-th component. 
The diffusion coefficients of BBHs from the $i$-th component are given by $D_i^{\rm NR}$, then 
the total diffusion coefficients of BBHs from all the background stars are given by
$D^{\rm NR}=\sum_i D_i^{\rm NR}$. 

For the RR processes, under the assumpsion that the process is a Gaussian random processes,
the contributions of multiple species on RR is approximately represented by a single 
species with effective mass given by $m_{\rm eff}=\langle m^2 \rangle(r)/\langle m \rangle(r)$, 
where the average is calculated at a given position $r$~\citep{2011MNRAS.412..187K}, the $\mu_M
=\sum \mu_{i,M}$, where $\mu_{i,M}$ is the procession due to the $i$-th component. 

We also consider the collision between the BBHs with the multiple species of background objects. 
The total collision rates are given by summing up the rates of individual background species, 
i.e., $R_{\rm tot}=\sum R_{i}$, where the rates of BBHs colliding with each species is given by $R_i$.

\item We now do not use the soft radius anymore in our explicit $3$-body simulation. Instead, 
 for both the BBH-star encounter and the BBH-MBH encounter we added the PN2.5 term to 
 consider the GW damping effects between any pairs of particles $1$ and $2$. 
 For particle $1$, these terms 
 are given by~\citep{Blanchet14} 
 \be
 \ba
 \vec{a}_{1,{\rm PN25}}&=\frac{4}{5c^5}\frac{G^2m_1m_2}{r_{12}^3}\left[
 \left(\frac{2Gm_1}{r_{12}}-\frac{8Gm_2}{r_{12}}-v_{12}^2\right)\vec{v}_{12}\right.\\
 &\left.+(\hat{r}_{12}\cdot \vec{v}_{12})
 \left(\frac{52Gm_2}{3r_{12}}-\frac{6Gm_1}{r_{12}}+3v_{12}^2\right)\hat{r}_{12}
 \right].
 \ea
 \ee
where $\vec{r}_{12}$ and $\vec{v}_{12}$ is the relative position and velocity vector between particle $1$ and 
$2$, respectively. $\hat{r}_{12}$ and $\hat{v}_{12}$ are the corresponding unit vectors.
The accelerations for particle $2$ can be obtained by changing the notion $1\leftrightarrow2$. 
$\vec{v}_{12}$  and $\hat{r}_{12}$ change sign by such an operation.

  We find that, by including such radiation, the merging rates of BBHs are increased by $\sim10\%$ compared to 
those in Paper I that is without such effect.

Including the GW damping effect can affect the dynamics of the 3-body encounters between a BBH and a star. 
For example, BBHs in the soft region can become `hard' after the encounter as the GW can radiative away some 
excess energy between the BBHs and the incoming star. This will also affect the 
outcome of the orbital corrections of the outer orbit mentioned below.
 
\item We added the orbital corrections on the outer orbit
of the BBH due to a fly-by event between the BBH and a background object.
We perform explicit 3-body simulation for each encounter and we can
obtain the change of relative velocity at infinity $\Delta \vec{v}'$ in the
center of mass frame of the three body before and after the encounter.
Denote $\vec{v}_b$  as the orbital velocity of the outer orbit of BBH
when the encounter happens, and $\Delta \vec{v}_b$ as its change due to the
encounter. Then simply we have
\begin{equation}
  \Delta  \vec{v}_b=\frac{m_3}{m_{\rm BBH}+m_3}\Delta  \vec{v}'
\end{equation}
where $m_3$ is the mass of the incoming background object. 

%{\ZFadd{The merge position of BBHs due to encounters with MBH: depend on the lifetime, if lifetime 
%is very short, then it is merged near the pericenter, if very long, then distributed uniformly in orbits.}}

Then the change of the orbital energy $\Delta E$ and angular momentum $\Delta \vec{h}$
in the MBH frame can be obtained according to corrections of this velocity, assuming that
the the encounter is located at position $|\vec{r}|=a_2$ of the orbit.
Such a correction introduce additional change in the energy and angular momentum of 
the outer orbit of a BBH, which is effectively an additional diffusion on the outer orbits of the BBHs, 
in additional to the two-body and resonant relaxation.  

\item Other minor refinements. For example, we refine the criteria of when a 
BBH is dominated by GW radiation during the phase of KL oscillation. A BBH is considered merging
only if $10|\dot{e}^{\rm KL}_1|<|\dot{e}^{\rm GW}_1|$ and $\delta t-t'<50 T_{\rm GW}$. 
If $|\dot{e}^{\rm KL}_1|>|\dot{e}^{\rm GW}_1|$, we then require $r_p<(M_A+M_B)G/c^2$, such that it
can be merged only if the horizon of each black hole touches with each other.

\end{enumerate}
\section{The GW harmonic}
\label{apx:GW_harmonic}
For the BBHs with chirp mass $\mathcal{M}=(m_Am_B)^{3/5}m_{\rm BBH}^{-1/5}$, and with 
none-zero eccentricity $e_1$, the energy usually emitted in a broad range of harmonics. 
The energy emitted per GW frequency at the $n$th harmonic at the 
rest frame of the source is given by~\citep[Eq. 20 of ][]{2015PhRvD..92f3010H},
\citep{2017MNRAS.470.1738C} 
\be
\frac{dE_n}{df_r}=\frac{(\pi G)^{2/3} (\mathcal{M})^{5/3}}{3(1+z)^{1/3}f^{1/3}}\left(\frac{2}{n}\right)^{2/3}
\frac{g(n,e_1)}{F(e_1)}
\ee
where function $g(n,e)$ is given by~\citep{Peters63}
\be
\ba
g(n,e_1)&=\frac{n^4}{32}\left\{\left[J_{n-2}(ne_1)-2eJ_{n-1}(ne_1)+\frac{2}{n}J_n(ne_1)
+2e_1J_{n+1}(ne_1)-J_{n+2}(ne_1)\right]^2\right.\\
&\left.+(1-e_1^2)\left[J_{n-2}(ne_1)-2J_n(ne_1)+J_{n+2}(ne_1)\right]^2+\frac{4}{3n^2}J_n(ne_1)^2\right\}
\label{eq:gne}
\ea
\ee
where $J_n$ are Bessel's functions
and $F(e_1)$ is given by 
\be
F(e_1)=\frac{1+\frac{73}{24}e_1^2+\frac{37}{96}e_1^4}{(1-e_1^2)^{7/2}}
\label{eq:fe}
\ee
The maximum number of harmonics necessary to a precision of $10^{-3}$ can 
be estimated as\citep{Oleary09}
\be
n_{\rm max}=5\frac{(1+e)^{1/2}}{(1-e)^{3/2}}
\ee
For simplicity here the maximum harmonics we considered is $\min(n_{\rm max},10^5)$.

The calculation of evolution of BBH ends at the last stable orbit (LSO). 
For a total mass of $m_{\rm BBH}$, the mean orbital frequency (assuming $e_1=0$) at LSO is given by 
 \be
 F_{\rm LSO}=\frac{c^3}{2\pi m_{\rm BBH} G}\frac{1}{6^{3/2}}
 \label{eq:lso}
 \ee
 
 \section{The drift of Fourier phase of BBHs}
 \label{apx:fourdrift}
Denote $t_\star$ as the proper time of BBH, $t_{\star,c}$ as the proper time of coalescence and
$l_c$ as the orbital phase before coalescence. If ignoring all relativistic effects, 
the time of arrival of the GW emitted from the BBHs to the observer is determined 
simply by the Reomer delay. Assuming Keplerian motion of BBHs around MBH, the time of arrival 
is given by 
\be
t =t_\star+ \Delta_{R}(t_\star),
\ee
where 
\be\ba
\Delta_R&=\frac{a_2}{c}\sin I_2\sin\omega_2(\cos E'-e_2)+\frac{a_2}{c}(1-e_2^2)^{1/2}\sin I_2\cos\omega_2\sin E',\\
&E'-e_2\sin E'=\frac{2\pi}{P_2}(t_\star-t_{\star,0}).
\ea\ee
Here $E'$ is the eccentric anomaly, $t_{\star,0}$ is the time of pericenter passage and $P_2$ is the 
orbital period of the outer orbit of BBH. $a_2$ and $e_2$ is the SMA and eccentricity of the outer orbit of the 
BBH, respectively. $I_2$  is the orbital inclination respect to the sky plane
and $\omega_2$ is the argument of periapsis of the orbit.

BBHs moving in a constant speed respect to the observer will cause a constant shift in measured chirp mass thus the line of signt velocity can not be 
reconstructed from observation. However, drift of the  observed phase of the GW
will be measurable if enough accelerations are presented during a period of observation. 
Similar to \citet{2017ApJ...834..200M}, we call the 
BBH moving around an MBH the ``source'' BBH, and suppose that there is a ``reference BBH'' 
that has the same velocity with the source BBH, i.e., $v_{\rm ref}$, at a given reference 
time $t_r$ in the observer's frame. For aLIGO/Einstein telescope, $t_{r}$ is usually the observed 
time of coalescence. For spaced-based telescopes, $t_{r}$ is either the time at the end of the mission, 
or the time when the characteristic strain of the BBH $h_c$ moves below the detection sensitivity. 
Denote $t_\star'$ and $t'$ as the proper time of BBH and time of arrival of the GW
 of the reference BBH, respectively, then 
 we can define the coordinate of time such that the proper time of both the source and the  
reference BBH at $t_r$ is both given by $t_{\star, r}$. Thus, in the reference BBH, we 
have 
\be
t'=t_\star'+\Delta_{R}(t_{\star,r})-\frac{v_{\rm ref}}{c}(t_{\star,r}-t_\star')
=t_\star'+\Delta_{R}(t_{\star,r})-\frac{v_{\rm ref}}{c}\left[t_{r}-\Delta_{R}(t_{\star,r})
-t_\star'\right]
\ee

Denote $l(t_\star)$ (or $l(t'_\star)$) as the orbital phase at proper time $t_\star$
of the source BBH (or $t'_\star$ of the reference BBH).
% then the phase difference between the source and the reference BBH at time $t$ of observer's frame is given by 
%\be\ba
%\Delta l&=l(t_\star)-l(t_\star')\\
%&=l[t-\Delta_R(t_\star)]-l\left[\frac{t-v_{\rm ref}t_r/c }{1+v_{\rm ref}/c}
%-\Delta_R(t_{\star,r})\right]
%\label{eq:deltapsinsmall}
%\ea\ee
For eccentric BBHs, the GW radiation has a broad range of harmonics. 
The Fourier transform of the GW can be calculated in 
the stationary phase approximation~\citep{1999amms.book.....B}. For 
a BBH with no relative motion respect to the observer, the Fourier
phase of the $n$-th harmonic at frequency $f$ is given by 
\be
\Psi_{n}(f)=-2\pi f t_\star(f)+n l[t_\star(f)]
\ee
where $t_\star$ is the solution of $ndl(t_\star)/dt_\star=f$.
For BBHs with $e<0.4$ the phase is provided by~\citet{2009PhRvD..80h4001Y} (ignoring other PN terms expect GW radiation)
\be\ba
\Psi_n(f)&=nl_c-2\pi ft_{\star,c}-\frac{3}{128x}\left(\frac{n}{2}\right)^{8/3}
\left[1-\frac{2355}{1462}e_i^2\chi^{-19/9}
+e_i^4\left(\frac{5222765}{998944}\chi^{-38/9}-\frac{2608555}{444448}\chi^{-19/9}\right)\right.\\
&\left.+e_i^6\left(-\frac{75356125}{3326976}\chi^{-19/3}-\frac{1326481225}{101334144}\chi^{-19/9}
+\frac{17355248095}{455518464}\chi^{-38/9}\right)\right.\\
&+e_i^8\left(-\frac{250408403375}{1011400704}\chi^{-19/3}+
\frac{4537813337273}{39444627456}\chi^{-76/9}-
\frac{6505217202575}{277250217984}\chi^{-19/9}\right.\\
&\left.\left.+\frac{128274289063885}{830865678336}\chi^{-38/9}\right)+
\mathcal{O}(e_i^{10})\right]
\label{eq:psin}
\ea
\ee
where $x=(\pi \mathcal{M}fG/c^{3})^{5/3}$ and $\chi=f/f_i$, $e_i$ and $f_i$ is the entering eccentricity and frequency of the BBH, respectively; $t_{\star,c}$ is the time of coalescence.

If considering the motion of the source BBH and those of the reference BBH, the 
Fourier phase difference between the two are given by
\be\ba
\delta \Psi_n (f)&=-2\pi f(t-t')-n[l(t_\star)-l(t_\star')]\\
&=-2\pi f(t-t')+\Psi_n(f_0)-\Psi_n(f_0')+2\pi (f_0t_\star-f_0't_\star')
\label{eq:deltapsin0}
\ea\ee
where
\be
f_0=n\frac{dl(t_\star)}{dt_\star}=\frac{f}{1+v(t_\star)/c}, ~~
f_0'=n\frac{dl(t_\star')}{dt_\star'}=\frac{f}{1+v_{\rm ref}/c}
\ee
%is the solution of $f_0=ndl(t_\star)/dt_\star$.

If $t_{\star}$ is very close to $t_{\star,r}$, denote $\delta t_\star=t_{\star,r}-t_{\star}$, 
using the fact that 
\be
\frac{d\Psi_n(f_0)}{df_0}=-2\pi t_\star
\label{eq:dpsitstar}
\ee
Then Equation~\ref{eq:deltapsin0} reduce to 
\be
\ba
\delta \Psi_n (\delta t_\star)
&\simeq \left(\frac{d\Psi_n(f_0)}{df_0}+2\pi t_\star\right)(f_0-f_0')
-\frac{2\pi f}{2c}a_{\rm ref}\delta t_{\star}^2\\
&\simeq-\pi f\frac{a_{\rm ref}}{c} \delta t_{\star}^2=-\pi n F_{\rm orb}(t_\star)\frac{a_{\rm ref}}{c} \delta t_{\star}^2
\label{eq:deltapsismall}
\ea
\ee
where $F_{\rm orb}$ is the orbital frequency of BBH, 
$a_{\rm ref}$ is the line of sight acceleration at proper time $t_{\star,r}$ 
(or $t_r$ at the observer's frame), which is given by 
\be
a_{\rm ref}=-\frac{\bh Gc}{r(t_{\star,r})^3}\Delta_R (t_{\star,r})
\ee
$r(t_\star)=a_2(1-e_2\cos E')$ is the distance to the MBH at local time $t_\star$ of the BBH.
Note that $t_{\star,r}$ depends on the order of harmonic $n$ as each harmonic rise above 
the noise level at different time of observation (for a given GW frequency $f$). 
If the reference time is the time of coalescence ($t_{\star,r}=t_c$), then
$\delta t_\star$ can be expressed as a function of frequency $f$, by using 
Equation~\ref{eq:dpsitstar}, up to orders of $e^{10}_i$:
\be\ba
\delta t_\star&=t_c-t_\star=
\frac{5}{256\pi x f}\left(\frac{n}{2}\right)^{8/3}\left[1-\frac{157}{43}e_i^2\chi^{-19/9}
+e_i^4\left(\frac{1044553}{56544}\chi^{-38/9}-\frac{521711}{39216}\chi^{-19/9}\right)\right.\\
&\left.+e_i^6\left(-\frac{15071225}{138624}\chi^{-19/3}-\frac{265296245}{8941248}\chi^{-19/9}
+\frac{3471049619}{25784064}\chi^{-38/9}\right)\right.\\
&+e_i^8\left(-\frac{50081680675}{42141696}\chi^{-19/3}+
%\frac{412941013691843}{591669411840}\chi^{-76/9}-
\frac{31764693360911}{45513031680}\chi^{-76/9}-
\frac{1301043440515}{24463254528}\chi^{-19/9}\right.\\
&\left.\left.+\frac{25654857812777}{47030132736}\chi^{-38/9}\right)+
\mathcal{O}(e_i^{10})
\right]
\label{eq:dtstar}
\ea\ee
Note that Equation~\ref{eq:psin} and~\ref{eq:dtstar} are only stable when $e_i\la0.4$.
 
 \section{Estimating GW backgrounds by Monte-Carlo integration}
 \label{apx:GWBK}
The multiple integration of Equation~\ref{eq:hcbkg_int} can be simply reduced to
a Monte-Carlo integration:
\be\ba
h_{c,bkg}^2(f)&\simeq\frac{4G^{5/3}}{3\pi^{1/3} c^2f^{4/3}}\frac{R_0}{N_{\rm tot}}
\sum_{i=1}^{N^{\rm tot}} \frac{ \mathcal{M}_i^{5/3}}{(1+z_i)^{4/3}}
\sum_{n=1}^\infty\left(\frac{2}{n}\right)^{2/3}\frac{g(n,e_{i})}{F(e_i)}\\
\ea\ee
where $N_{\rm tot}$ is the size of the MC samples (usually $10^5$).
$\mathcal{M}_i$, $z_i$ and $e_i$ is the chirp mass, redshift and the eccentricity of the 
$i-$th Monte-Carlo samples. $R_{0}$ is the total merging rate given by 
\be\ba
R_0&=\iiint \frac{dN(\mathcal{M},e_0,z)}{dzde_0d\mathcal{M}} dz de_0 d\mathcal{M}\\
&=\iiiint\frac{dN(\mathcal{M},e_0,z |\bh)}{dt_r de_0d\mathcal{M}} \frac{dt_r}{dz}
n(\bh,z) dz de_0 d\mathcal{M} d\bh\\
&=\iint \frac{R(\bh) n(\bh, z)}{(1+z)H_0E(z)}d\bh dz\\
\ea\ee
where $E(z)=[\Omega_M(1+z)^3+\Omega_\Lambda]^{1/2}$, $R(\bh)$ is the merging rate 
in the local universe for galaxies with a given MBH mass
(Equation~\ref{eq:RS}), $t_r$ is local time of BBH,
$n(\bh,z)$ is the number distribution function of MBH given by~\citet{2015ApJ...810...74A}.
The integration is over all redshift and MBH mass ranges.

According to~\citet{2017MNRAS.470.1738C}, this summation can be further reduced to 
\be\ba
h_{c,bkg}^2(f)\simeq&\frac{R_0}{N_{\rm tot}}
\sum_{i=1}^{N^{\rm tot}} \left(\frac{\mathcal{M}_i}{\mathcal{M}_0}\right)^{5/3}
\left(\frac{1+z_i}{1+z_0}\right)^{-1/3}\\
\times& h_{c,0}^2\left(\left.f\frac{f_{p,0}}{f_{p,t}}\right|e_0, f_0\right)\left(\frac{f_{p,t}}{f_{p,0}}\right)^{-4/3}\\
\label{eq:hcbkg_monte_carlo}
\ea
\ee
where $\mathcal{M}_0=4.16\times10^8\msun$, $z_0=0.02$. 
$h_{c,0}^2(f|e_0, f_0)$ is given by~\citet{2017MNRAS.470.1738C} in unit of Mpc$^{-3}$, 
and $f_{p,0}$ is given by the following equation when $f=0.1$nHz and $e=0.9$:
\be
\frac{f_p}{f}=\frac{1293}{181}\left[\frac{e^{12/19}}{1-e^2}(1+\frac{121e^2}{304})^{870/2299}\right]^{3/2},
\ee
$f_{p,t}$ is given by the above equation when $f=f_0$ and $e=e_0$.

To generate Monte-Carlo samples, we take the following steps: 
\begin{enumerate}
\item Draw a total number of $N_{\rm tot}=10^5$ random pairs of $\bh, z$ following
the two-dimensional distribution:
\be
P(\bh, z)\propto \frac{R(\bh) n(\bh, z)}{(1+z)H_0E(z)}
\ee
\item For the $i$-th mass and redshift from above (deonted by $M_{\bullet,i}$ and $z_i$, respectively), get the samples from each model that is with the mass of MBH 
closest to $M_{\bullet, i}$ (one of the four {{\texttt{ GNC}}} runs with $\bh=10^5\msun, 10^6\msun, 10^7\msun$ and $10^8\msun$). Randomly select one of the sample and get its corresponding value of 
$\mathcal{M}_i$ and $e_{0,i}$.
\item Calculate the GWB according to Equation~\ref{eq:hcbkg_monte_carlo} given the 
$\mathcal{M}_i$, $e_{0,i}, z_i$, $i=1, \cdots, N_{\rm tot}$ of the MC samples for each value of frequency $f$. 
\end{enumerate}

\end{document}